\documentclass[twocolumn, amsmath, amssymb, aps, physrev]{revtex4-2}

\usepackage{graphicx}
\usepackage{dcolumn}

\usepackage{accents}

\usepackage{placeins}

\newcommand*{\dt}[1]{%
  \accentset{\mbox{\large\bfseries .}}{#1}}
  
\usepackage{lineno}

\usepackage{amsmath} 
\usepackage{amssymb}
\usepackage{fancyhdr} 
\usepackage{geometry} 
\usepackage{xr-hyper}

\newcommand*{\addFileDependency}[1]{
\typeout{(#1)}
\IfFileExists{#1}{}{\typeout{No file #1.}}
}
\makeatother

\usepackage{mathtools}

\usepackage{orcidlink}

\makeatletter

\makeatletter

\usepackage{hyperref}

\begin{document}

\title{Task learning through stimulation-induced plasticity in neural networks}

\author{Francesco Borra\orcidlink{0000-0002-2410-3448}}
\author{Simona Cocco\orcidlink{0000-0002-4459-0204}}
\author{Rémi Monasson\orcidlink{0000-0002-1852-7789}}
\affiliation{Laboratoire de Physique de l'Ecole Normale Sup\'erieure, PSL Research, CNRS UMR8023, Sorbonne Universit\'e, 24 rue Lhomond, 75005 Paris, France%
 }


\begin{abstract}
Synaptic plasticity dynamically shapes the connectivity of neural systems and is key to learning processes in the brain. To what extent the mechanisms of plasticity can be exploited to drive a neural network and make it perform some kind of computational task remains unclear. This question, relevant in a bioengineering context,  can be formulated as a control problem on a high-dimensional system with strongly constrained and non-linear dynamics. We present a self-contained procedure which, through appropriate spatio-temporal stimulations of the neurons, is able to drive rate-based neural networks with arbitrary initial connectivity towards a desired functional state. We illustrate our approach on two different computational tasks: a non-linear association between multiple input stimulations and  activity patterns (representing digit images), and the construction of a continuous attractor encoding a collective variable in a neural population. Our work thus provides a proof of principle for emerging paradigms of {\em in vitro} computation based on real neurons. 
\end{abstract}

\maketitle

\section*{Introduction}

Natural neural networks have long been a key source of inspiration for machine learning and computing, starting from Rosenblatt's perceptron to neuromorphic computing~\cite{furber2016large} and to nowadays deep learning. Recently, it was suggested that biological neural structures might be directly exploited as a support for {\em vitro} computation~\cite{organoidintell,kagan2022vitro}. Progress over the last decade has made it possible to  grow, preserve, and study brain organoids~\cite{orga1}. The capabilities of stimulating and recording neural populations~\cite{deisseroth2011optogenetics, yuste2018, shin2021} allow for interfacing these systems, potentially turning brain organoids into miniature biological computers. Computing with organized assemblies of neurons could offer considerable advantages with respect to electronics-based devices, in particular in terms of low energy consumption, massive parallel computation, and continuous learning. However, besides the practical challenges raised by these technologies~\cite{qian19}, it remains  unclear how they could be best used to achieve high-level computation. The aim of this paper is to address one conceptual question arising in this bio-engineering context: how could circuits of biological neurons be trained to carry out desired computational tasks? 

Computation with neural cultures has been largely based so far on the framework of reservoir computing~\cite{tanaka2019recent, sumi2023biological}, in which the readout of the high-dimensional neural activity is trained to perform the task of interest. However, reservoir computing falls short from fully realizing the computational potential of neural systems, as it does not exploit their capability for reconfiguration of biological connectivity across time. The presence of plasticity mechanisms, involving multiple molecular and cellular processes, is crucial to the formation and maintenance of experience--related changes to neural function and circuitry~\cite{citri}. These mechanisms can potentially be exploited to shape, through appropriate stimulations, the activity and the connectivity of networks~\cite{carrillo}, as has been shown in specific settings~\cite{liu2023creation}. From a bio-engineering point of view, the question is therefore to determine spatio-temporal stimulation patterns capable of remodeling a plastic neural network and make it achieve a desired computational task (Fig.~\ref{fig:fig1}(a)). Establishing guiding principles and practical tools to obtain those stimulation patterns is essential to future progress in biological computation.

From a mathematical point of view, supervised learning of a task is generally cast as an optimization problem in very high dimensions. Given a population of $N$ (artificial) neurons connected through $S$ synaptic connections $J_{ij}$, 
one looks for the minimum of the cost (loss) function $U\big(\{J_{ij}\}\big)$ 
expressing the mismatch between the target and actual computation carried out by the network. In artificial nets, the cost $U$ 
can be gradually reduced through gradient descent (or one of its stochastic variants) in the space of connections, until a minimum with good performance is reached. A key point here is that all moves in the $N^2$-dimensional interaction space, in particular the one along the gradient of the cost, $\frac{\partial U}{\partial J_{ij}}$, 
are allowed during learning. 

The situation is much more constrained in the case of biological networks, in which the learning dynamics  can obviously not be arbitrarily chosen, and plasticity mechanisms set strong limitations about feasible directions. As a concrete example, consider the case of Hebbian-like plasticity rules, in which the changes in the connections $\Delta J_{ij}$ 
are functions of the firing rates $r_i$ (over some appropriate time scale) 
of the neurons. As  the number $N$ of neurons is generally much smaller than the number $N^2$ of synaptic interactions, the plastic changes $\Delta J_{ij}$ 
are highly inter-dependent, and cannot be individually tuned to match the gradient components $\frac{\partial U}{\partial J_{ij}}$. 
This fundamental limitation makes supervised learning with biologically-plausible rules conceptually much more intricate than with unconstrained dynamics. In addition, the magnitude of synaptic modifications is hard to control in biological networks, while the capability to implement adaptive learning rates is generally regarded as a key ingredient in machine learning.

In the present paper, we show that {\em in silico} neural networks that obey plausible plasticity mechanisms can be effectively trained, through the application of adequate control stimulations, to carry out very diverse computational tasks. These stimulations vary with time and from neuron to neuron, and are obtained by solving a sequence of optimization problems. Intuitively, the control stimulations are
capable of inducing adequate neural activity, progressively driving the network connectivity through its intrinsic Hebbian-like plasticity towards a desired state  (Fig.~\ref{fig:fig1}(a)).
Figure~\ref{fig:fig1}(b) summarizes our learning protocol, which relies on multiple cycles of stimulations and recordings of the neural population. At the beginning of each cycle, the responses of the network to few short probing stimulations are recorded, and used to infer the connectivity of the network. Based on this estimate of the connectivity, we plan a control stimulation pattern, which is subsequently applied to the neurons. Under this control, plastic changes to the connections take place and enhance the network performance in achieving the desired computation. The procedure is iterated until the computational target is reached.

\section*{Model}

\subsection*{Dynamics of neural activity}

We  consider a network of $N$ neurons, characterized by their firing rates $\mathbf{r}(t)=\{r_i(t)\}$ at time $t$ and the time-dependent synaptic connectivity matrix $\mathbf{J}(t)=\{J_{ij}(t)\}$ (Fig.~\ref{fig:fig2}). By convention, $J_{ij}$ refers to the coupling from the pre-synaptic neuron $j$ to  the post-synaptic neuron $i$. The neural population includes $N_E$ excitatory ($E$) and $N_I$ inhibitory ($I$) neurons, which constrains the signs of the corresponding synaptic interactions. The activities of the neurons obey the standard dynamical rate equations
\begin{equation}\label{phi11}
\tau_n \, \frac{d r_i}{dt} (t)=-r_i(t)+\Phi\left(\sum_{j} J_{ij}(t)\, r_j(t)+f_i(t)\right)
\end{equation} 
where $\tau_n$ is the membrane relaxation time, and $f_i(t)$ is a time-dependent control stimulation on neuron $i$, which can be dynamically controlled at will in the range $[f_{min};f_{max}]$. The input-to-rate transfer function $\Phi$ is a sigmoidal function, ranging between zero and maximal frequency $r_{max}$ (Methods and SI, Section~\ref{supp:model}).
Control stimulations $f_i$ are expressed in units of $r_{max}$, while connections $J_{ij}$ are dimensionless  (SI, Section~\ref{supp:model}). 

\subsection*{Synaptic plasticity}\label{supp:stationary}

Plasticity induces activity-dependent changes in the interactions. While the precise rules describing  plasticity are still debated in neuroscience, we use a simple and general model presenting three essential qualitative features: Hebbian associativity, regression towards a baseline (homeostasis),  and bounded synaptic strengths. 
The equation for the synaptic dynamics is
\begin{multline}\label{eqpl}
\tau_s \;\frac{d J_{ij}}{dt} (t) = \underbrace{\eta(\epsilon_j)\; (r_{i}-\theta(\epsilon_j))\;r_{j}}_{\mbox{hebbian}}\\
-\underbrace{\beta_1\;|J_{ij}|\;(r_{i}^2-\theta_0(\epsilon_j)^2)}_{\mbox{homeostasis 1}}\\
-\underbrace{\beta_2\; \text{sign}(J_{ij})\; h_2\big(|J_{ij}|-\bar J\big)}_{\mbox{homeostasis 2}} \,
\end{multline}
In the equation above, $\epsilon_j=E$ or $I$ is the pre-synaptic neuron type, and $h_2(u)=u^2$ if $u\ge 0$, 0 if $u<0$. The first term assumes that changes in the connections derive from a Hebbian-like covariance rule with a post-synaptic threshold depending on the neuron type \cite{vico2003stable}. The last two terms account for homeostatic feedback~\cite{turrigiano2008self,turrigiano2000hebb,zenke2017temporal,zierenberg2018homeostatic}, biasing the activity of the post-synaptic neuron towards a baseline activity $\theta_0$, and imposing a soft clipping of synaptic amplitudes outside the range $[-\bar J;+\bar J]$. 
Plasticity for excitatory and inhibitory connections are assumed to have different learning rates $\eta(E)$ and $\eta(I)$ and are associated to different thresholds $\theta(E)$ and $\theta(I))$ to enhance the stability of the network activity states. The parameters $\beta_{1}$ and $\beta_2$ control the strengths of the homeostatic constraints. 

The learning rule in Eq.~\eqref{eqpl} is flexible, and can describe Hebbian as well as anti-Hebbian learning. The former is obtained when  $\eta(\epsilon_j)$ is positive, independently of the  pre-synaptic neuron type $\epsilon_j$. To accommodate for anti-Hebbian inhibition \cite{foldiak90}, one chooses $\eta(E)>0, \eta(I)<0$. We report results with both choices of rules below. We stress that anti-Hebbian here refers to a negative feedback  between correlations and synaptic strength change. 

While the associative learning rule above can be easily modified, a key assumption we rely on is that synaptic changes are slow compared to the fast variations in the activity, {\em i.e.} $\tau_n \ll \tau_s$ (Methods and Supplemental Material (SM), Section~\ref{supp:stationarity}). Due to this assumption only the slow variations (on the $\tau_s$ time scale) of the stimulations $\mathbf{f}(t)=\{f_i(t)\}$ matter. Furthermore, the neural activities $r_i$ are quasi-stationary and locked to the slowly varying inputs. In mathematical terms, the r.h.s. of Eq.~\eqref{phi11} vanishes.  

\subsection*{Target state of the network}

\begin{widetext}
\begin{figure*}[h!]
    \centering
     \includegraphics[width=\textwidth]{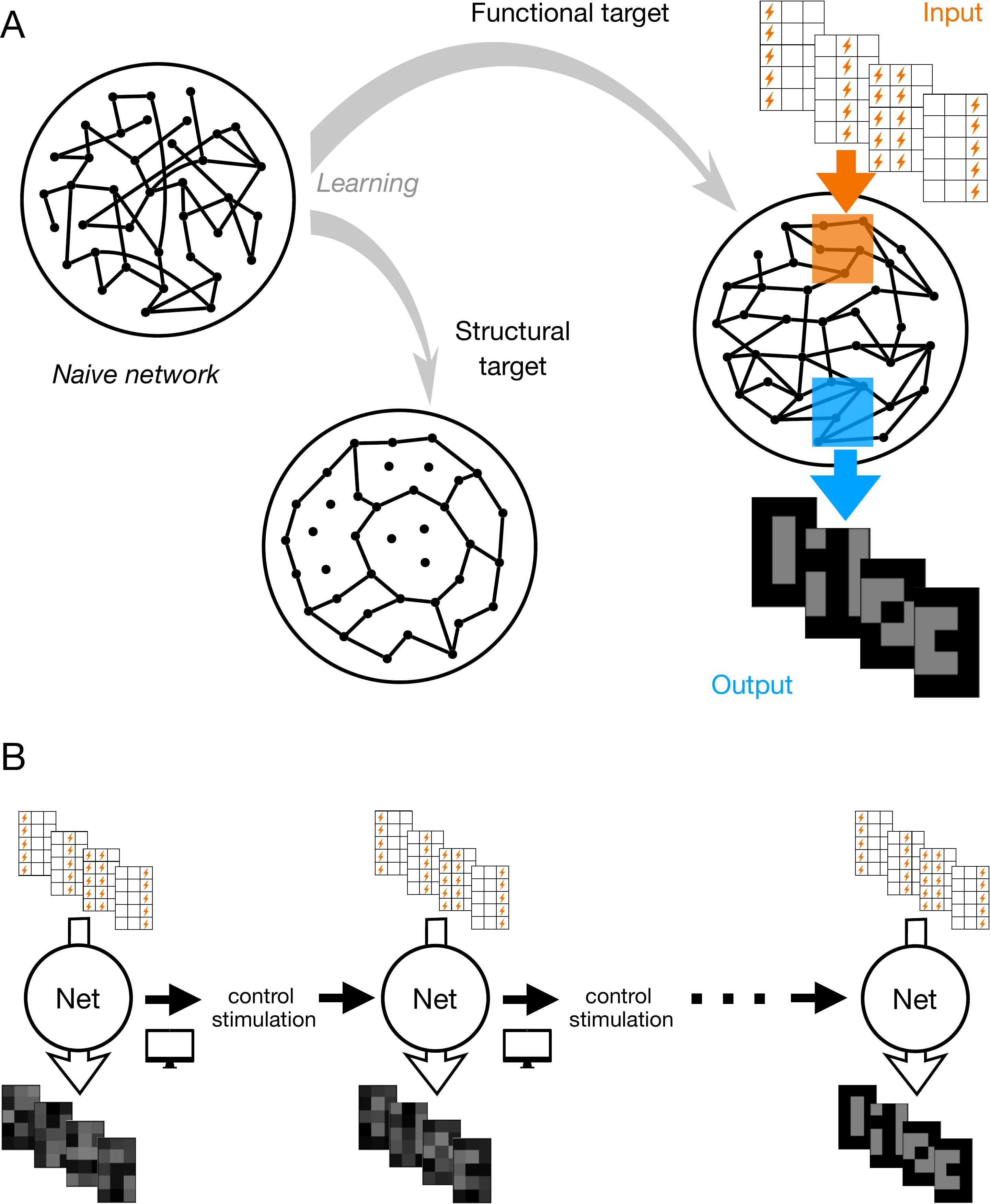}
    \caption{\fontsize{8pt}{11pt}\selectfont
{\bf Computational targets for plastic neural networks.} \\
{\bf (a)} Our goal is to reconfigure a naive neural network and reach some target, either structural (a specific connectivity state) or functional (for instance, the network is required to implement some input--output associations). This reshaping is achieved through a learning process, in which appropriate spatio-temporal stimulations of the neurons exploit intrinsic plasticity mechanisms.\\
{\bf (b)} The computational target, here, the functional task in panel (a), is reached through a learning cycle. Prior to the learning process, the naive network associates the inputs to random outputs (left). Our algorithm computes the best control to be applied to the network to modify its connectivity through plastic changes. Applying this control stimulation to the neural population results in an enhancement of the network performance (middle).  The control is then re-optimized and applied during a new stimulation period. After multiple iterations, the correct input-to-output association is achieved (right).
    }
    \label{fig:fig1}
\end{figure*}
\end{widetext}

Our goal is to train the neural network connectivity $\mathbf{J}$ to meet some target, either structural or functional (Fig.~\ref{fig:fig1}(a)). In the former case the network connectivity is asked to reach some  value, say, $\mathbf{J}^{target}$.  The training procedure should ensure that the loss  
\begin{equation}\label{cost_struct}
    U_{task}\big(\mathbf{J}\big) =  \sum_{i,j }w_{\epsilon_i,\epsilon_j}\, \big[ J_{ij}-J^{target}_{ij} \big]^2 
\end{equation}
decays towards zero over the training time. We introduce weights $w_{\epsilon_i,\epsilon_j}$ depending on the neuron types to  ensure that the four classes of connections $E,I\to E,I$ are  equally contributing to $U_{task}$ (Methods). 

In the functional case, the target is not the connectivity itself as above, but the computation carried out by the network. The neurons $i$ in the network are partitioned into three subpopulations, referred to as input ($in$), processing ($proc$), and output ($out$). The network is required to implement a set of $n_{pairs}$ input/output mappings, that is, produce desired  activities $r_i^\mu$ of the neurons $i$ in the $out$ subpopulation in response to specific input stimulations $\mathbf{f}^\mu$ applied to the neurons $j$ in the $in$ subpopulation, with $\mu=1,...,n_{pairs}$. A possible loss  associated to this association task  reads
\begin{equation}\label{cost:functional}
    U_{task}\big(\mathbf{J}\big)  = \sum_{\mu=1}^{n_{pairs}}\sum_{i\ \in\ out}\left[ r_i\big( \mathbf{J},\mathbf{f}^\mu \big) -r_i^\mu\right]^2 \ ,
\end{equation}
where $r_i(\mathbf{J},\mathbf{f})$ refers to the stationary solution of Eq.~\eqref{phi11}; see Methods for alternative choices of the loss. The task may be made harder by requiring that the $in$ and $out$ neurons  share no direct connections, and communicate through the neurons in the $proc$ subnetwork. 

\section*{Training loop: supervised control of Hebbian plasticity}

\begin{widetext}
\begin{figure*}[h!]    
    \centering
     \includegraphics[width=.9\textwidth]{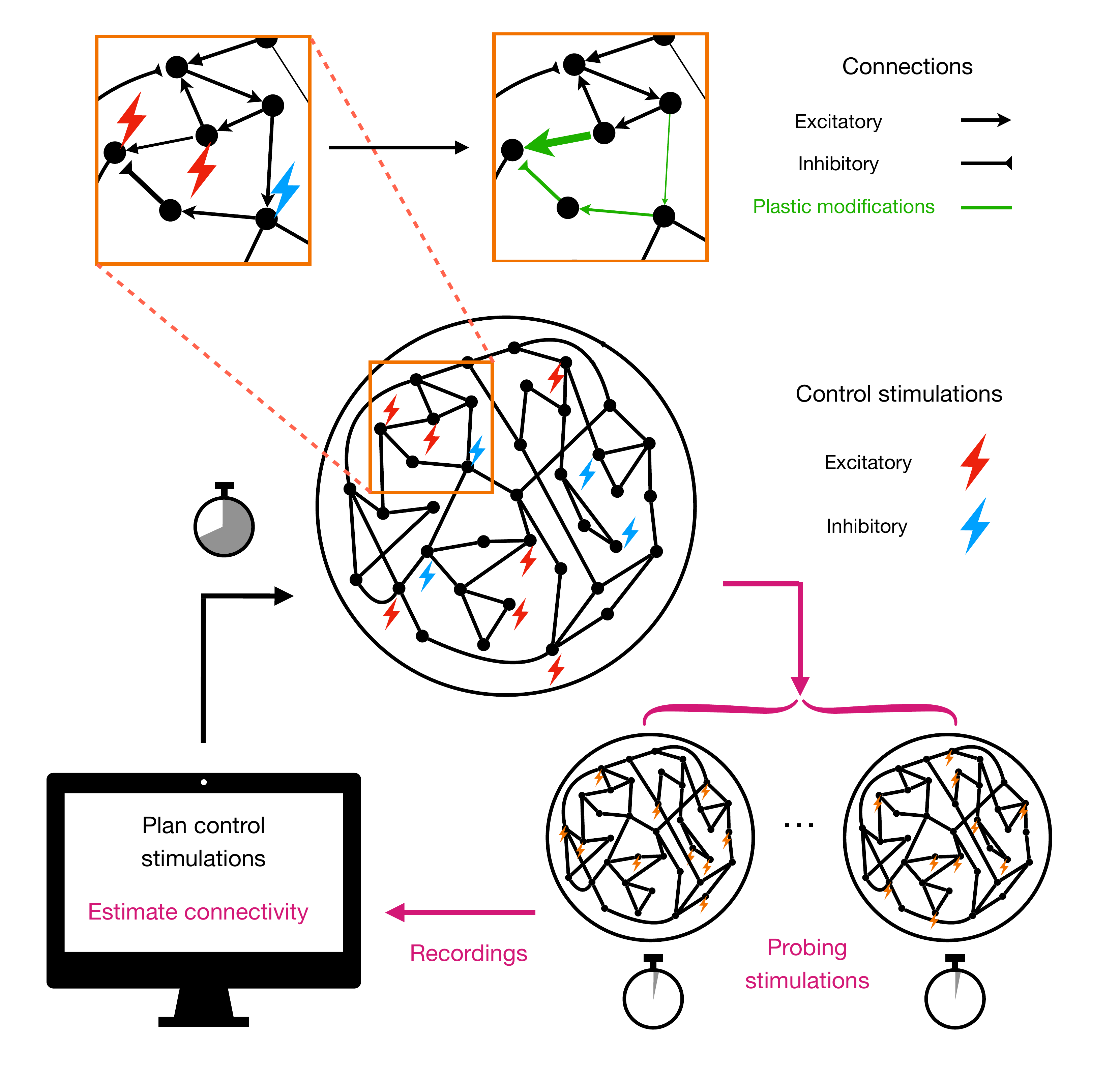}
    \caption{\fontsize{8pt}{11pt}\selectfont{\bf  Stages of the learning cycle: planning of control stimulations, reconfiguration under plasticity effects, and connectivity estimation}\\
In each cycle, a control stimulation $f_i^*$ is applied to the neurons $i$ in the network, generating stationary firing rates $r_i$ (middle). In turn this activity pattern leads, through the plasticity rule, to specific strengthening or weakening of the connections $J_{ij}$, indicated by the changes in the thicknesses of the connection arrows (top). Once the control period halts, few short and random stimulations are applied to the neurons, and the corresponding activities are recorded (bottom, right). These data are then used to update the estimate of the network connectivity, and to plan the optimal control for the next cycle (bottom, left).
}
    \label{fig:fig2}
\end{figure*}
\end{widetext}

Our training procedure is based on a sequence of learning cycles, labelled by $k=0,1,2,...$ and sketched in Fig.~\ref{fig:fig2}. Each cycle $k$ can be decomposed in three steps:
\begin{itemize}
\item Estimation of the current connectivity, $\mathbf{J}_k$, through fast probing of the responses of the network to random stimuli.
\item Calculation of the optimal control to be applied, $\mathbf{f}_k^*$, to shift the network connectivity state towards the desired target.
\item Application of this control during the period $\Delta t$, leading to a reorganization of the network through its intrinsic plasticity mechanisms.
\end{itemize}
The learning process stops when the value of the loss $U_k$ is considered small enough that the target has been reached. 
We now give more details about the steps above; technicalities can be found in Methods. 

\subsection*{Estimation of the network connectivity}

Determining and updating the optimal control stimulation  requires knowing, to some degree of accuracy, the state of  connectivity $\mathbf{J}_k$~\cite{de2018connectivity}. To mimic realistic conditions, $\mathbf{J}_k$ cannot be accessed through direct measurements. We therefore infer the connectivity based on multiple recordings of the activity (Fig.~\ref{fig:fig2}). We probe the network stationary activity states $\mathbf{r}^\nu=\{r_i^\nu\}$ corresponding to various random stimulation patterns $\mathbf{f}^\nu=\{f_i^\nu\}$, where $\nu=1,...,n_{probes}$. The connectivity matrix is then estimated so that the stationary equations
\begin{equation}\label{eq:stat}
   r_i^\nu =  \Phi \bigg( \sum_{j} (J_k)_{ij}\, r_j^\nu+f_i^\nu \bigg)
\end{equation}
are fulfilled for all $i=1,...,N$ and $\nu=1,..., n_{probes}$, see Eq.~\eqref{phi11} and Methods. Increasing the number $n_{probes}$ of recordings reduces the error over the estimated connectivity matrix, but the duration of the whole process should be sufficiently short not to induce any plastic modification to $\mathbf{J}_k$: hence $n_{probes}$ should not exceed $\tau_s/\tau_r$,where $\tau_r$ is the relaxation time of the network activity, which can be estimated from $\tau_n$ and ${\bf J}_k$ (Methods). In practice, the connectivity varies little after each control stimulation period, and few recordings are needed to update our estimate.

\subsection*{Calculation of optimal control}
\begin{widetext}
\begin{figure*}[h!]
    \centering
\includegraphics[width=.7\textwidth]{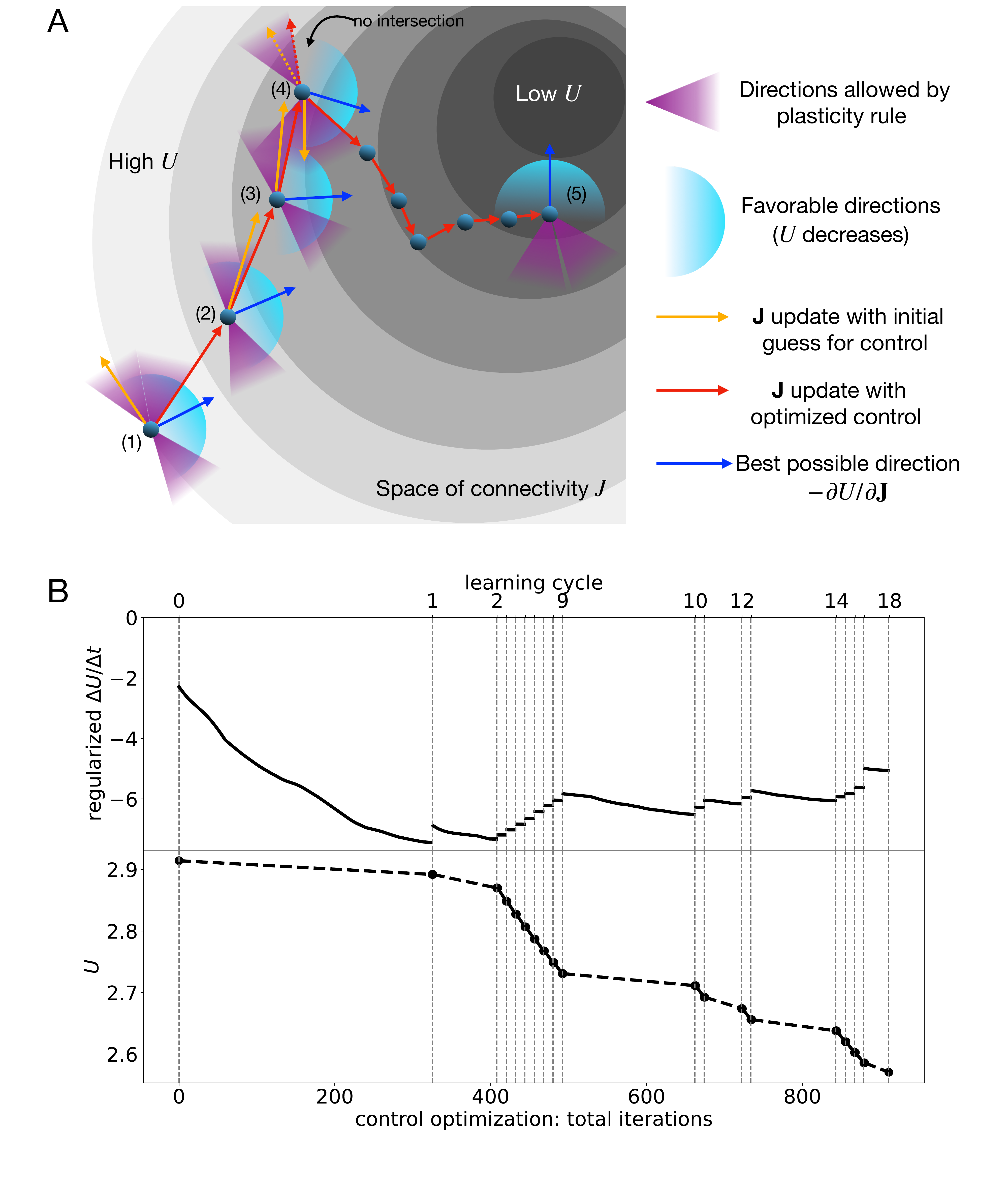}
    \caption{\fontsize{8pt}{11pt}\selectfont  
  {\bf  Schematic dynamics of the network and feasible directions in the high-dimensional connectivity space during the training process.} \\
  {\bf (a)} Grey levels show values of the cost $U$, which quantifies the mismatch between the current connectivity $\bf{J}$ of the network and the target.  Dark blue balls locate the connectivity $\mathbf{J}$ at the beginning of each control stimulation period. The set of all directions in the $N^2$-dimensional connection space along which $U$ decreases is shown by the cyan area, centered around the direction of steepest descent of $U$ indicated by the blue arrows. The purple cone symbolically represents the set of all feasible directions for $\mathbf{J}$ under the plasticity dynamics, {\em i.e.} which can be reached under any $N$-dimensional control stimulation $\mathbf{f}$. \\
{\bf (1)} Initial situation, prior to any control stimulation. The synaptic change corresponding to a random $\mathbf{f}$ may point to a `bad' direction outside the cyan region (orange arrow), while the change associated to the optimal control stimulation $\mathbf{f}^*$ lies on the edge of the purple cone (red arrow), as close as possible to the best direction (blue arrow).  \\
{\bf (2)} After one control stimulation period. Repeating the same control stimulation as before would lead to a suboptimal change of $\mathbf{J}$ (orange arrow), while the best  stimulation (red arrow) yields a larger decrease in $U$. \\
{\bf (3)} Control stimulations are updated to ensure optimal synaptic changes  until {\bf (4)} all local updates of the previous $\mathbf{f}^*$ lie outside the `good' (cyan) region. We then choose a new initial $\mathbf{f}$, and resumes the search for the optimal $\mathbf{f}^*$ as in cycle (1). This process is iterated, until the change in $U$ are very small and the optimization is completed, or the feasible and `good' regions do not overlap any longer (point {\bf (5)}). The performance of the network (final value of $U$) is then assessed. \\
{\bf (b)} Decay rate of the loss during the optimization over the control (top) and resulting loss $U_k$ after each learning cycle $k$ (bottom), for  the setting corresponding to Task 1. Abscissa indicates the number  $n$ of optimization steps. Vertical dashed lines locate the beginnings of the training cycles $k$.
Top: decay rate $(U(n)-U_k)/\Delta t$ at step $n$ of the optimization algorithm, calculated with the regularized loss, see Eq.~\eqref{deltaU}. For each cycle, the rate is decreased through gradient descent over $\mathbf{f}$ until a plateau is reached. The optimization then halts, and the control defines $\mathbf{f}^*_k$. At the begining of the new cycle, the network connectivity $\mathbf{J}$ has changed, the previous control is not optimal any longer, and the optimization process resumes. The plateau values of the rate increase over cycles, making progressively harder to find a good control to train the network.  
         }
    \label{fig:fig3}
\end{figure*}
\end{widetext}

Ideally, one would like the change of connectivity during one cycle, $\Delta{\mathbf{J}}=\mathbf{J}_{k+1}-\mathbf{J}_k$,  to align along the direction of steepest descent of the loss $U$, i.e. $\Delta{\mathbf{J}}\propto -  \partial U/\partial\mathbf{J}|_{\mathbf{J}_k}$
However, the connections obey the plasticity rule in Eq.~\eqref{eqpl} and their dynamics is not directly and arbitrarily controllable. From an informal point of view, the best we can hope for is to find a stimulation that will enhance neuronal activities in such a way as to move connections in a direction as aligned as possible with the gradient of $U$, see Fig.~\ref{fig:fig3}(a).


Let us call $\Delta \mathbf{J}(\mathbf{f},\Delta t, \mathbf{J}_k)$ the change in the connections produced by applying the stimulation $\mathbf{f}$ for a period of time $\Delta t$ to the network with initial connectivity state $\mathbf{J}_k$. In principle, $\Delta \mathbf{J}$ is obtained by integrating the Hebbian-like plasticity rule in Eq.~\eqref{eqpl} during the time interval (under fixed stimulation). Then, the best control is formally given by
\begin{equation}\label{eqdecU}
\mathbf{f}^*_k = \underset{\mathbf{f}}{\text{argmin}} \quad 
U\big(\mathbf{J}_k+\Delta\mathbf{J}(\mathbf{f},\Delta t, \mathbf{J}_k)\big) 
\ .
\end{equation}
Solving this optimization problem and finding the absolute minimum can be quite hard. We use a gradient descent procedure in the space of the controls $\mathbf{f}$, not to be confused with the training dynamics over $\mathbf{J}$, which is guaranteed to return a local minimum at least. We stress that this optimization step is an abstract computation, unrelated to any physical or biological process, and is done on a computer separate from the neural model, see Fig.~\ref{fig:fig2}. Informally speaking, each cycle can be thought of as an approximate gradient descent of $U$ in the $\mathbf{J}$-space, whose precise direction is determined by another gradient descent in the $\mathbf{f}$-space. This double process is shown in Fig.~\ref{fig:fig3}(b).


The determination of $\mathbf{f}^*_k$ in Eq.~\eqref{eqdecU} via gradient descent is a non trivial computational problem due to the non-linearities in the neural and synaptic dynamics, see Methods. Choosing $\Delta t\ll\tau_s$ (SM,~Section~1) and imposing additional regularization terms (Methods) ensure that the changes  $\Delta\mathbf{J}$ remain small at the end of the learning cycle.

\subsection*{Plastic reorganization of the network under control stimulations}

Once the optimal control $\mathbf{f}^*_k$ has been computed (in a time we neglect compared to the time scales at play in the neural network), it is applied to the neurons for a period $\Delta t$. The activities of the neurons rapidly adapt to these external stimulations, and the connections start evolving under the plasticity dynamical rules.

At the end of this training cycle, the network connectivity may differ from the expected value  $\mathbf{J}_k+\Delta\mathbf{J}(\mathbf{f}^*_k,\Delta t, \mathbf{J}_k)$ due two main factors. First, the estimate $\mathbf{J}_k$, obtained prior to the stimulation, is not equal to the ground truth connectivity. The accuracy depends on the number of measurements, and is limited by the constraint of fast probing as explained above. Second, the true underlying plastic mechanisms controling the network evolution can never be exactly modeled. The unavoidable mismatch between the network dynamics and Eq.~\eqref{eqpl} induces, at the end of the stimulation period, a systematic bias in the value of the connectivity. To account for this bias and quantify its consequences, we consider below two variants of Eq.~\eqref{eqpl}, with different set of parameters, for the `true' plastic evolution of the network and for its model counterpart used to compute $\Delta \mathbf{J}(\mathbf{f},\Delta t, \mathbf{J})$ and the optimal control $\mathbf{f}^*$.

\section*{Applications}

We apply below our learning procedure to two computational tasks; a third task defined on a small network is detailed in SM, Section~\ref{supp:task0}, and in the Discussion section.

%
%

\subsection*{Task 1: Non-linear association between all-or-nothing inputs and digit-like outputs}\label{subsec:task1}

\paragraph*{Definition of the task.} A fundamental computation carried out by neural circuits in organisms is the association of diverse behavioural (or motor) responses to multiple sensory inputs. This task is particularly non trivial to learn when it is non additive, as it cannot be realized by linear networks \cite{xor}, and may require to recruit dedicated brain areas \cite{bees}.

\begin{figure*}[h!]
    \centering
     \includegraphics[width=1
\textwidth]{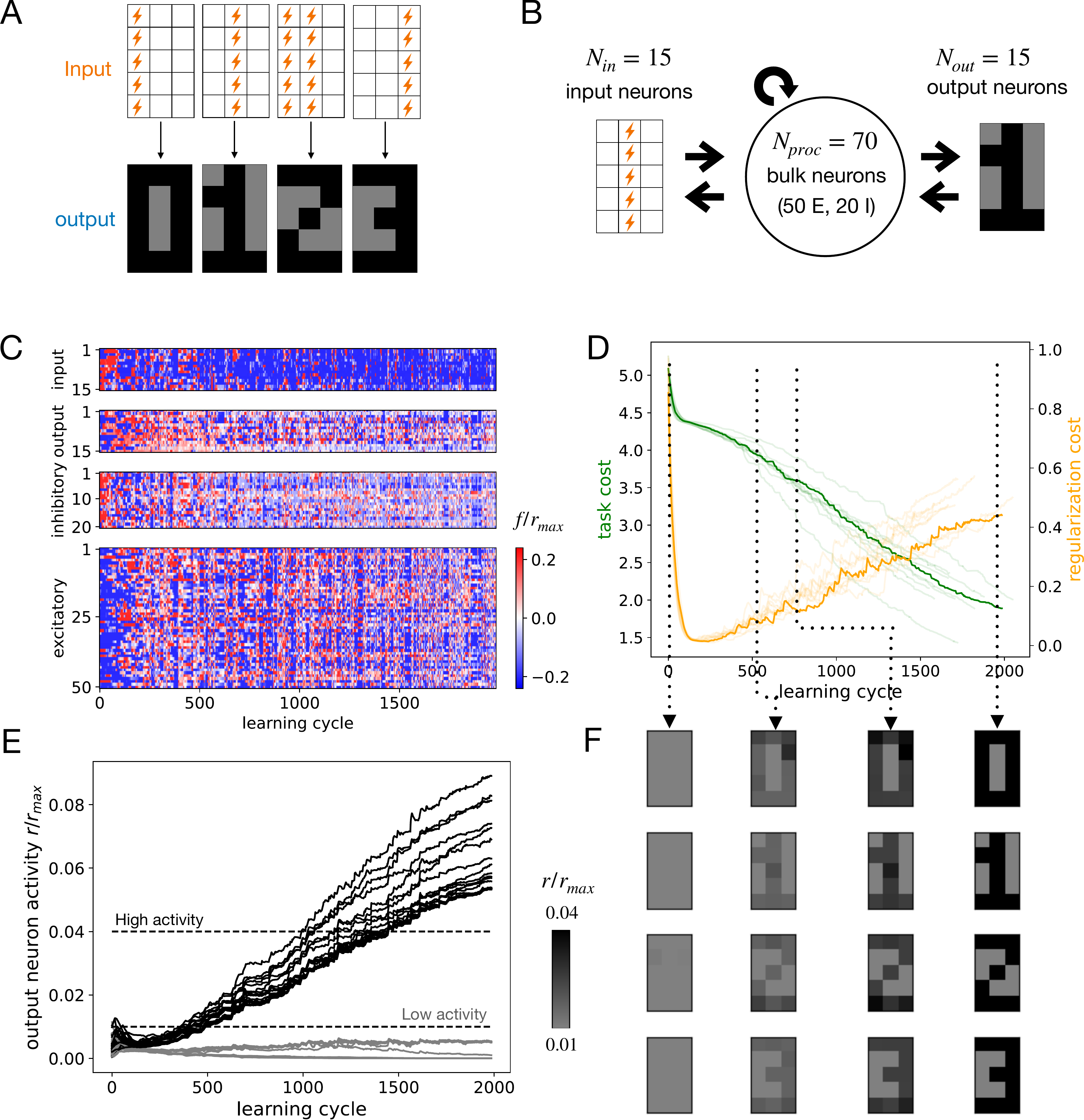}
    \caption {    \fontsize{8pt}{11pt}\selectfont
   {\bf Task 1: non-additive input-output association.}\\
{\bf (a)} The $n_{pairs}=4$ pairs of input-output associations to be learned by the network. Input patterns: flashes locate the input neurons (out of $N_{in}=15$) subject to a strong stimulus ($f_i=f_{max}$), while the other neurons (empty squares) receive no input ($f_i=0$). Output patterns: target activities of the $N_{out}=15$ output neurons (black: $r_i>0.04 r_{max}$, light gray: $r_i<0.01 r_{max}$). \\
{\bf (b)} Sketch of the network, with  the bulk processing area including $N_{proc}=70$ neurons, connecting the input and output neurons. The overall fractions of excitatory and inhibitory neurons in the network are equal to, respectively, 80\% and 20\%.\\
{\bf (c)} Control stimulations $f_i(t)$ (see color bar for values) applied to the neurons as a function of the learning step. From top to bottom: input, output, bulk inhibitory, and bulk excitatory neurons. \\
{\bf (d)} Costs associated with the task (green, left scale) and with regularization (orange, right scale) as functions of the learning cycle. \\
{\bf (e)} Activities of the output neurons in response to the input stimulation patterns during learning. The color (black or light gray) of each one of the $n_{pairs}\times N_{out}=60$ curves indicates the level of activity requested for the corresponding output neuron (high or low, see panel (a)). Training is successful when the black and light gray curves become well separated.\\
{\bf (f)}
Average activities of the output neurons for the four time steps identified with dashed line in panel (d), showing how the figures of digit emerge from the initial random image. Black and light gray activities are consistent with panel (e), with intermediate gray levels defined in the bar.
    }
    \label{fig:fig4}
\end{figure*}

We consider here a toy version of a non-linear associative task, in which a subset of $N_{in}$ input neurons can be stimulated in $n_{pairs}$ distinct ways. Each input stimulation is expected to elicit an associated activity pattern over $N_{out}$ output neurons, see Fig.~\ref{fig:fig4}(a). For the sake of visual clarity, the output patterns are represented as digits, with grey/black pixels corresponding to neurons with low/high activities arranged on a rectangular grid. We make sure that
\begin{itemize}
   \item the association task is  non-linear by carefully choosing the input/output pairs. For instance, the sums of the input stimulations associated to the output digits 0 and 1 is the input required to elicit digit 2; however,  2 is  quite distinct from the superimposition of $0$ and $1$, see Fig.~\ref{fig:fig4}(a). As a result of non-linearity the association task cannot be simply implemented through separate neural pathways, activated by one input and silent for the other ones.
    \item  there is no direct connection from the input to the output neurons, see Fig.~\ref{fig:fig4}(b). This  arrangement mimics the idealized structures of  a modular chip. Hence, the information about the input must be processed and conveyed by an intermediate processing ($proc$) region of the network, which contains most of the $N$ neurons.  This requirement also ensures that the decrease of the cost function in the initial part of the training is not due to a simple creation of a linear direct input-to-output connection, but rather to the creation of a complex structure.
\end{itemize}

\paragraph*{Training: costs and dynamics.} To learn the task, we introduce a cost over the connectivity matrices favoring the desired input/output associations, see Methods, Eq.~\eqref{maincosttask}. An additional cost aims at regularizing the connectivity and speeding up convergence towards stationary states of activity, see Methods.

We start from a network with weak, random connections. As our spatio-temporal control stimulation is applied to the neurons (Fig.~\ref{fig:fig4}(c)), the connectivity gets modified, and the costs associated to the task and to the regularization decrease, as shown in Fig. \ref{fig:fig4}(d). The activities of neurons progressively cluster into two categories, depending on their expected values in the output patterns (Fig.~\ref{fig:fig4}(e)). The learning process is characterized by three stages. First, the activities of all output neurons approximately reach the same low activity level independently of the input pattern, as regularization tends to decrease the amplitudes of the connections and of the activities. Then activities start to separate according to the output patterns.   At this stage one can already guess the output digits, see Fig.~\ref{fig:fig4}(f). We let the protocol proceed until the two groups (active, inactive) are clearly separated, and the output neuron activities match the target patterns, compare Fig.~\ref{fig:fig4}(a)\&(f).

\begin{figure*}[h!]
    \centering
     \includegraphics[width=.8
\textwidth]{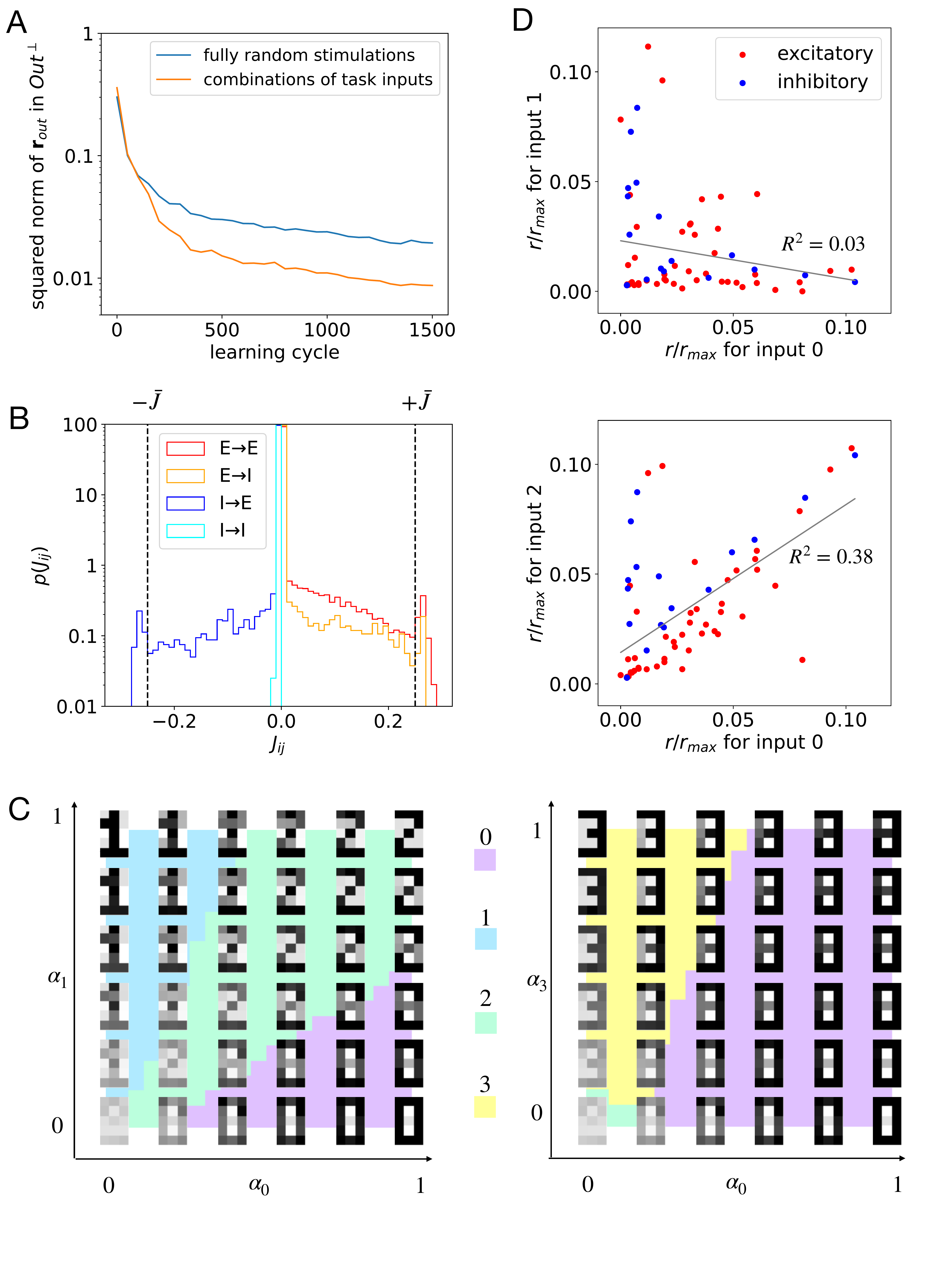}
    \caption {\fontsize{8pt}{11pt}\selectfont
     {\bf Learning dynamics and internal representations for Task 1.}\\
{\bf (a)} Squared norm of the projection of the output activity (over $N_{out}=15$ neurons) orthogonal to the space spanned by the $n_{pairs}=4$ digit patterns as a function of the number of training cycles. The total activity is normalized to unity. Blue curve: case of random 0-1 stimulations over the $N_{in}=15$ input neurons; orange: stimulations are random convex combinations of the $n_{pairs}$ inputs. Results were averaged over 10 training trials, and each point is averaged over 50 random input stimulations.\\
{\bf (b)} Histograms of interactions $J_{ij}$ between the neurons in the $E$ and $I$ populations. Dashed lines show the soft bounds in $\pm \bar J$.\\
{\bf (c)} Output neuron activities (gray levels) in response to combinations of inputs $\alpha_\mu \, \bf{f}_\mu+\alpha_\nu\, \bf{f}_\nu$, with $0\le \alpha_\mu,\alpha_\nu\le 1$. The bottom left corner corresponds to vanishing input and the bottom, right and top,left corners to pure inputs associated to, respectively, digits $\mu$ and $\nu$. The background colors (middle squares) indicate the most resembling digits. (Left) case $\mu=0,\nu=1$: the output of the linear combination of inputs is not the linear combination of the outputs, as required by the association task. (Right) Case $\mu=0,\nu=3$: the response to linear combinations of ${\bf{f}}_0$ and ${\bf{f}}_3$ is approximately additive.\\
{\bf (d)} Scatter plot of the activities of the 70 bulk neurons for inputs $0,1$ (top) and $0,2$ (bottom) after training. The representations associated to patterns 0 and 1 are not correlated, as expected from the orthogonality of ${\bf{f}}_0$ and ${\bf{f}}_1$  (Fig.~\ref{fig:fig4}(a)). Conversely,  the representations of  0 and 2 are strongly correlated, reflecting the relation ${\bf{f}}_2={\bf{f}}_0+{\bf{f}}_1$ (Fig.~\ref{fig:fig4}(a)). 
    }
    \label{fig:fig5}
\end{figure*}

\paragraph*{Transfer of information and neural representations.}
How the information is transferred from the input to the output neurons and represented in the $proc$ area is investigated in Fig.~\ref{fig:fig5}. 
While the input and output sub-populations in our network both include $N_{in}=N_{out}=15$ neurons, the associative task effectively takes place in a lower-dimensional space with $n_{pairs}=4$ dimensions. We show in Fig.~\ref{fig:fig5}(a) that learning progressively concentrates most of the $N_{out}=15$-dimensional activity in the $4$-dimensional subspace spanned by the output digits. This dimensionality reduction phenomenon is even stronger when the $N_{in}$ input neurons are stimulated by random combinations of the training inputs associated to the digits, implying the effective creation of a 4-to-4 dimensional channel upon learning. From a circuit point of view, this functional channel is supported by a network of connections, whose histogram is reported in Fig.~\ref{fig:fig5}(b). We observe that, during learning, the E and I subpopulations strengthen their connections, with the exception of recurrent interactions within the I neurons that remain weak. The creation of input-specific sub-networks of activity during the learning process can be seen in the animation provided in SI, Section~\ref{SI:gif1}.

We stress that, within this low-dimensional space, the network computation is not linear as imposed by the non-additivity of the task. We show in Fig.~\ref{fig:fig5}(c) how the network response to linear combinations of the inputs may largely differ from the superimposition of the responses to the single inputs, see case $0-1$ (left). This result confirms that the network is capable of learning complex behaviour beyond the superimposition of patterns in the low-dimensional subspace spans by the patterns. On the contrary, in the absence of any specific constraint imposed by the task, computation appears to be approximately linear, see combinations of inputs 0 and 3 producing the 8 digit (right).

Furthermore, we may ask how the network, after training, is able to  process and differentiate the patterns outside the input and output sub-populations. We see in Fig.~\ref{fig:fig5}(d) that inputs produce the activation of a significant fraction of the neurons in the $proc$ area and that many of those neurons respond specifically to distinct inputs, with stronger correlations in the evoked activities for more correlated inputs. In particular, the role of inhibitory neurons is key to the differentiation of input 2 from both 0 and 1, and to depart from additivity as imposed by the task. Further information about how neurons in the {\em proc} region participates in the association task can be found in SM, Fig.~\ref{figS:figNC}.

\subsection*{Task 2: Ring-like connectivity supporting continuous attractor dynamics}

\begin{figure*}[h!]
    \centering
     \includegraphics[width=.9\textwidth]{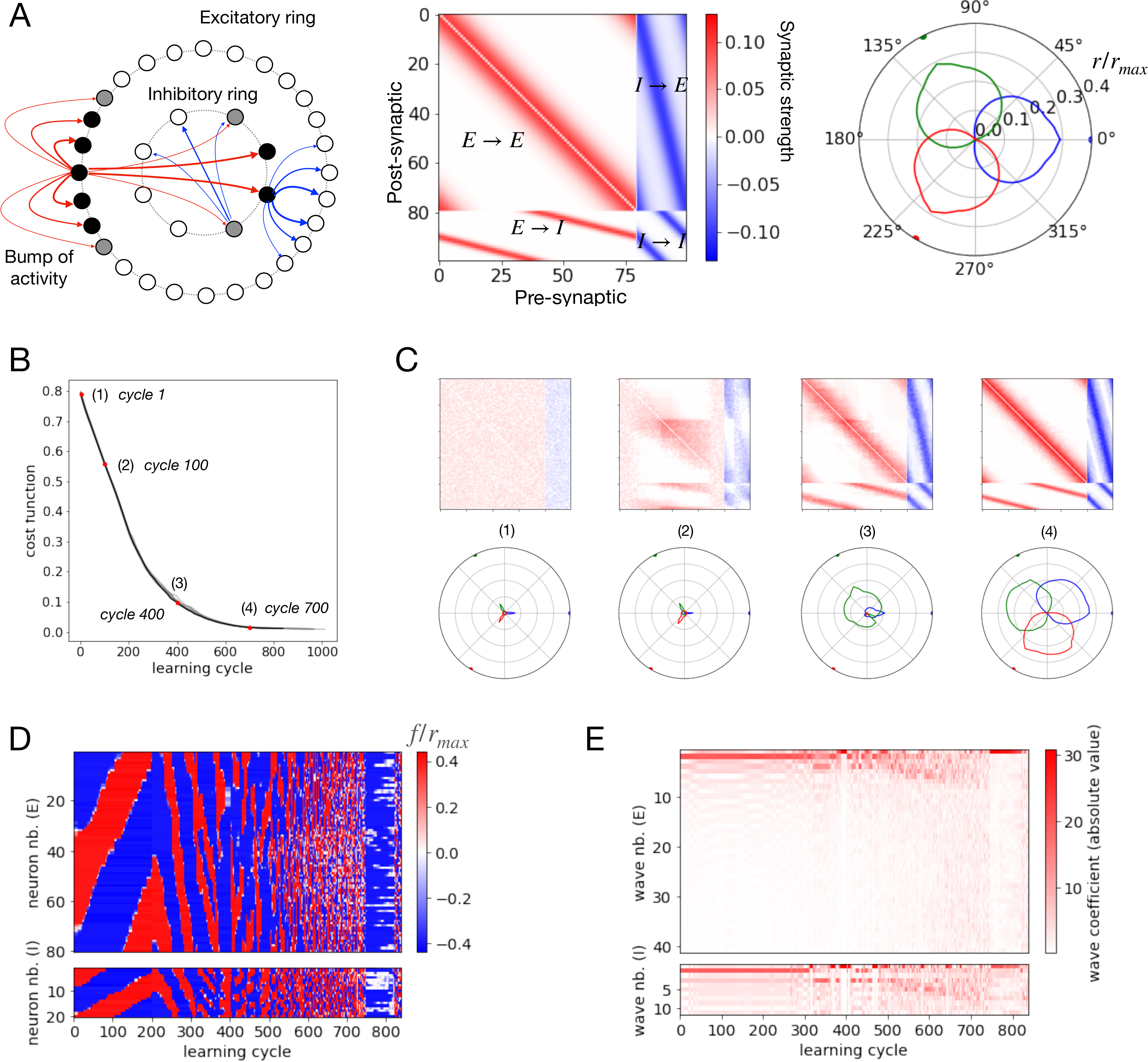}
    \caption{\fontsize{8pt}{11pt}\selectfont
  {\bf Task 2: building a continuous attractor.} \\
{\bf (a)} Target network (left) and connectivity matrix $\mathbf{J}^{target}$ (middle). Excitatory neurons (E) are arranged on the outer ring ($N_E=80$), and inhibitory neurons (I) on the inner one ($N_I=20$). Neurons on the E ring have strong excitatory connections with their neighbours, and project to inhibitory neurons diametrically opposed on the I ring. Inhibitory neurons repress neighbouring neurons on the E ring and repress the opposite side of the I ring, leading to a localization of the activity (bump). Right: Radial plots of receptive fields of neurons associated to angles 0 (blue), 120 (green) and 240 (red) degrees on the E ring. A weak localized input stimulation is applied for each angle (Methods) and the activities (averaged over 100 trainings) of all neurons in the stationary state are shown.\\
{\bf (b)} Task cost evolution during learning for 10 random naive networks (gray curves); the back line shows the average cost. Four representative learning cycles, labelled (1), (2), (3), (4) are referred to in panel (c).\\
{\bf (c)} Connectivity matrices (top) and receptive fields (bottom) at cycles (1), (2), (3), (4), showing different stages of learning. Same color codes as in panel (a).\\
{\bf (d)} Control stimulations $f_i(t)$ as functions of the learning cycle.\\
{\bf (e)} Amplitude $\hat f_k^\epsilon(t)=\sum_{\ell=1}^{N_\epsilon} f_\ell (t) \,
\cos(2\pi  k\,\ell/N_\epsilon)$ of the Fourier modes $k=0,1,...,N_\epsilon/2$ associated to the control stimulations shown in panel (d), for excitatory ($\epsilon=E$) and inhibitory ($\epsilon=I$) neurons. Initially, the control stimulation mainly consists of large waves (low-$k$ Fourier modes), while at the end of  training, modes at large $k$ are used to refine the synaptic structure on short angular scales. Notice the global suppression at cycle $\simeq 750$, after the receptive fields are formed.
    }
    \label{fig:fig6}
\end{figure*}

\paragraph*{Definition of the task.} We now aim at reshaping network connectivity into a target matrix $\mathbf{J}^{target}$ through learning. The target connectivity defines a ring attractor, capable of supporting a bump of activity coding for a continuously varying angle. Ring attractors were first theoretically hypothesized \cite{amari1977,benyishai1995}, and have recently been observed in the ellipsoid body of fly \cite{jayaraman}. Neuron connections are organized along two rings, see Fig.~\ref{fig:fig6}(a). Excitatory neurons form the outer ring, with connections decays with their distance, while inhibitory neurons compose the smaller inner ring. The connections between the two rings are such that a bump of activity on one side of the outer ring induces inhibition on the diametrically opposite side (see Fig.~\ref{fig:fig6}(a)).  The position of the bump is arbitrary in the absence of external input, and is otherwise attracted by a weak and localised input to the outer neurons.

Control stimulations are computed with the cost function given by Eq.~\eqref{cost_struct}. The time behaviour of the cost is shown in Fig.~\ref{fig:fig6}(b) for ten different random initial networks (see Fig.~\ref{fig:fig6}(c), left, for a realization of the naive connectivity). After a fast initial decay the cost relaxes to very low values, signalling the success of the learning procedure. 

\paragraph*{Training: network connectivity and stimulations.}
Snapshots of the connectivity at different steps of the learning process are displayed in Fig.~\ref{fig:fig6}(c). The strong similarity between the network connectivity at the end of learning (compare Figs.~\ref{fig:fig6}(a) and (c), step (4)) confers the desired functional properties of continuous attractors to the network. We report in Fig.~\ref{fig:fig6}(d) the average firing activities of neurons when a weak input stimulation (intended to pin the bump of activity at a specific angle) is applied.  At the initial state of learning (1), the network responds  by a weak excess activity simply reflecting the localized pinning input stimulation. Through training, a sub-population of neurons emerges, whose activity is supported by the recurrent connections. This sub-population varies with the angle associated with the input stimulation, see (2) and (3). The emergence of receptive fields is improved with further training cycles (4). Notice that, even when the target structure is very well reconstructed, the direction of the polarization is not perfect, due to tiny correlated defects in the connectivity structure. The presence of these defects does not preclude the accuracy of the coding of angular information in the network \cite{kuehn2023}.

The time behavior of the control stimulations applied to neurons during learning are shown in Fig.~\ref{fig:fig6}(e). We observe the presence of long-distance (large angle) correlations at small times, which disappear at the end of the training. To better characterize this angular structure, we plot the Fourier transform (along neuron indices) of the control stimulations in Fig.~\ref{fig:fig6}(f). In the initial stage of training, the stimulation is essentially represented by cosine and sine waves, with periods matching the E and I ring extensions. As learning proceeds, higher and higher wave number modes are recruited to refine the short scale structure of the synaptic matrix. This observation is in agreement with the progressive emergence of connectivity in Fig.~\ref{fig:fig6}(c).

%
%

\section*{Discussion}

We have shown how appropriate stimulations can induce plastic reconfiguration in a network subject to Hebbian-like plasticity processes and make it capable of accomplishing a prescribed computation. Our approach could successfully be applied to  different tasks (see also SM, Section~\ref{supp:task0}) and  plasticity rules (Hebbian and anti-Hebbian, see SM, Sections~\ref{aht2} and~\ref{supp:ht3}). These results can be thus seen as a proof-of-concept for general-purpose neural training, which would compile a task into a neural network by externally guiding the intrinsic learning mechanisms. At its core, our training method addresses and solves a sort of inverse Hebbian problem: instead of looking for how synaptic interactions change under external inputs, we search for an appropriate set of inputs, or control stimulations implementing desired changes in the interactions through plasticity mechanisms. Notice that, strictly speaking, the issue of modifying  connectivity  with constrained control is not unique to  biological neural networks:  neuromorphic circuits, e.g. based on memristors can undergo plastic changes~\cite{serrano2013stdp}, and computation in such networks could be among potential applications of our training protocol. 
We now discuss our results, both from the computational and the bioengineering/biological points of view.

\subsection*{Computational aspects}

\paragraph*{Constrained learning and approximate gradient descent.} As stressed in the introduction, plasticity rules strongly constrain the manifold of possible changes in the connectivity matrix and straightforward gradient descent of the loss is generally not feasible (Fig.~\ref{fig:fig3}(a)). This fundamental limitation is clearly present even in the case of very simple tasks trainable on small networks (Supplemental Material (SM), Section~\ref{supp:task0} and SM, Fig.~7). As the number $N$ of neurons increases, the ratio between the number of control variables ($N$) and the dimension of the connectivity space ($N^2$, or a fraction of $N^2$ depending on network sparsity) diminishes, and the situation ought to become less and less favorable. Fortunately, this effect is counterbalanced by the huge increase in the multiplicity of the networks realizing the task with $N$~\cite{Marder}. We observe, indeed, that Task~1 is easier to learn when the size $N_{proc}$ of the processing area (Fig.~\ref{fig:fig4}(b)) increases, 
see SM, Section~\ref{supp:param} and SM, Fig.~\ref{figS:figF}. As a result, functional targets are easier to reach than structural ones, as the latter cannot benefit from any multiplicity of solutions. 

\noindent
An additional difficulty is the impossibility to use an adaptive learning rate, a key ingredient in the convergence of most machine-learning algorithms. Though some variability in the duration $\Delta t$ of the control can be considered, this is constrained by biological and experimental time scales and cannot vary over orders of magnitude. As a result, the magnitude of the changes $\Delta \mathbf{J}\propto \Delta t \;\dot{\mathbf{J}} $ in the connectivity after each cycle cannot be tuned at will.

\paragraph*{Choice of cost function: structural vs functional targets.}
The nature of the cost $U$ has also a deep impact on the time course of performance throughout the learning process. Any task  could, in principle, be given a structural cost, {\em e.g.} by running gradient descent of the functional cost on a computer and finding an adequate network $\mathbf{J}^*$ implementing the target task. However, in the structural case, the value of $U$ is not immediately informative about the computation carried out by the network. Good functional performance, such as the ability to create angle-specific receptive fields in Task~2, see Fig.~\ref{fig:fig6}(d), may be reached for vanishing $U$ only (SM, Section~\ref{aht2} and SM, Fig.~\ref{figS:fig2anti}(a)\&(b)), which is difficult to achieve in practice. For this reason, functional costs, for which low values of $U$ directly imply good performances, are preferable. Accordingly, we find that defining $U$ from the worst input-output mismatch, see Methods, Eq.~\eqref{maincosttask}, rather than from the average over all pairs as in Eq.~\eqref{cost:functional}, gives better performance for Task~1. 

\noindent
A consequence of the multiplicity of solutions in the functional case is that relatively little changes to the initial (and random) network are generally sufficient to meet the target, compared to what is needed to meet a structural target. This intuition is corroborated by the analysis reported in SM, Section~\ref{supp:Corr} and by the animation provided in SM, Section~\ref{SI:gif2}. We observe that the internal activity of the network remains strongly self-correlated in time during function-based training, providing evidence for the ability of the learning procedure to reach a functional target with moderate changes to the connectivity.

\paragraph*{Model uncertainty and robustness.}
A crucial point to be addressed in future works is the impact of noise and uncertainty. Dynamical noise, both in the neural activity and in the synaptic transmission,  could be taken into account through a probabilistic formulation of the control problem. In addition, one should consider measurement errors in the activity recordings, affecting the estimation of the connectivity, or in the stimulation process, leading to inaccurate control of the network. 

\noindent Taking into account model uncertainty is more delicate. As a matter of fact, neither the equations describing the neural activity dynamics, nor the plasticity rules are exactly known. Any inaccuracy in the model is potentially a source of bias and of correlated errors in the determination of the stimulation protocol. To estimate the impact of such errors, we have studied the performance of our learning scheme in a mismatch setting, in which the plasticity rules governing the evolution of the connectivity and the ones used for the computation of the protocol differed. We show in SM, Section \ref{SI:error}, that successful training is achieved even with 10$\%$  (relative) mismatches on the parameters entering the plasticity rules. 
This result is a good indication of the robustness of our approach to uncertainties in the model. Needless to say, controlling errors in the modelling of the plasticity rules and designing robust stimulation schemes will be crucial for future developments.

\paragraph*{Importance of the range of controls.}
At the implementation level, our training procedure involves high--dimensional optimization over the control stimulation $\mathbf{f}$, and requires careful numerical implementation. A number of parameters in the optimization loop have to be set, whose choice impacts not only the outcome but also the speed of learning, see SM, Sections~\ref{supp:task1}\&\ref{supp:task2}. Of crucial importance is the range of variation of the control stimulation applicable to the neurons, $f_{min}\le f_i\le f_{max}$. If the range is wide, a lot of variability in the control stimulation schemes is possible, and learning is easy and fast. As $f_{max}-f_{min}$ decreases, the cost decays more slowly (SM, Fig.~\ref{figS:figF}(a)). Further investigations would be needed to better understand if a minimal value of $f_{max}-f_{min}$ exists below which the target task is not reachable any longer and, more generally, how reachability depends on the task. Notice that the protocol we have implemented corresponds to a control strategy with zero time horizon, which optimizes the cost change $\Delta U$ at each cycle.  Global protocols, derived in the framework of optimal control~\cite{kirk2004optimal}, could possibly achieve higher  performance. However, our high-dimensional setting with incomplete knowledge (of connectivity) could suffer from serious error propagation over large time horizons, making implementation  difficult.

\paragraph*{Continual learning.}
The final configuration of the network (reached at the end of the control scheme) is in general not a stable state of the learning dynamics and would be lost over time. This is a general issue with systems undergoing continual learning \cite{contlearn}. From a computational point of view, one would either need to re-apply stimulations to maintain the neural circuit in its state or use a cost function that ensures that the final configuration is stable. From a biological point of view, plasticity processes can be downregulated in many ways {\em in vivo} \cite{citri}. Diminishing plasticity is for instance essential at certain stages of brain development \cite{dudek}.

\subsection*{Biological aspects.}


\paragraph*{A new framework,  {\em in vitro} biological computing.}

Our work is deeply motivated by the emerging field of {\em in vitro} computation with biological neurons. 
The main approach in this domain so far has been  reservoir computing~\cite{lukovsevivcius2009reservoir,george2014input,tanaka2019recent,yada2021physical,sumi2023biological}, which exploits the capability of randomly connected neural populations to produce highly variable patterns of activity, and has been linked to plausible neurobiological mechanisms~\cite{rigotti2010internal,enel2016reservoir}. 
Associative effects of plasticity can be exploited for the sake of learning, and reach performance beyond standard reservoir setups~\cite{morales2021unveiling,cai2023brain}. More targeted approaches have been proposed to directly exploit plasticity for task-training, for instance by choosing familiar/unfamiliar inputs as rewards/punishment~\cite{kagan2022vitro} or by exploiting stimulation avoidance~\cite{masumori2015emergence}.  Our approach, on the contrary, aims at directly controlling  and training the neural structure through the determination of adequate stimulations to be imposed to the neurons. 

\noindent
The idea of learning through stimulations is not new to biology, and is the basis for the concept of vaccination. The immune system is endowed with learning mechanisms that allow it to respond to pathogens after appropriate stimulations, e.g. following the administration of a pathogen protein or its coding RNA sequence. The vaccination scheme (timing and dosage) is crucial to the quality of the immunization~\cite{molari}, and can be designed to enhance the production of the so-called broadly neutralizing antibodies having large response spectra~\cite{bnabs}. By analogy, one may expect that stimulation protocols optimized to the learning dynamics of  neural systems could drive them to desired operational states.


\paragraph*{Limitations and plausibility.}
Our modelling approach is highly simplified and unlikely to capture detailed biological mechanisms, and is primarily intended to explore conceptual challenges in the training of networks with constrained plasticity. Though the learning procedure is robust against mismatches between the 'true' and modelled plasticity mechanisms (SM, Section \ref{SI:error}), it relies on a number of assumptions that need to be discussed in terms of their technological and biological plausibility. First, we have assumed that stimulations could be done at the level of individual neurons: fine stimulation and recording capabilities are being developed~\cite{adesnik2021probing} but, as of today, optogenetic~\cite{deisseroth2011optogenetics,stamatakis2018simultaneous} and electrophysiological~\cite{spira2013multi} techniques rather allow one to target groups of nearby neurons. This limitation could be taken into account in our approach by imposing spatial correlations between the components of $\mathbf{f}$. Second, our protocol  could be adapted to constrained controls $\mathbf{f}$, {\em e.g.} whose components have a given sign (though excitatory and inhibitory tools are available~\cite{land2014optogenetic,han2007multiple,deisseroth2011optogenetics}), or discrete (on/off) rather than continuous values. Though these limitations could in principle be enforced, it remains to be seen how they would affect  the learning procedure, both in terms of performances and training time.

\paragraph*{Separation of neural and plasticity time scales.} An important hypothesis is that the time scale $\tau_r$ associated to the relaxation of neural activity is much shorted than the duration of the control stimulation during a cycle, $\Delta t$, over which the connectivity changes. Under this assumption, the neural activity is stationary, and  the estimate of the connectivity can be updated by  probing the network response to few random stimuli without inducing synaptic changes (Methods, Eq.~\eqref{eqJupdate}). Physiological time scales related to single-neuron dynamics ($\tau_n$ in Eq.~\eqref{phi11}) are expected to be of the order of tens of milliseconds. It is then crucial that network effects do not slow down the relaxation of the population activity on times $\tau_r\gg \tau_n$, see Methods, Eq.~\eqref{defd}. This is ensured in practice in Task~1 through the introduction of a regularization cost on the connectivity, $U_{reg}$, which prevents the emergence of slow dynamics modes in the network (Methods, Eq.~\eqref{ureg}). We have checked that $\tau_r$ remains comparable to $\tau_n$ throughout the learning process for all the tasks considered here (SM, Section~\ref{supp:stationarity}). While it is difficult to estimate  with precision the time scale $\tau_s$ associated to plasticity effects in Eq.~\eqref{eqpl} and, consequently, the duration $\Delta t$ of control stimulation, a biologically plausible order of magnitude for $\Delta t$ is of a few tens of seconds (SM, Section~\ref{supp:model}). Note that this is loosely consistent with meaningful plastic modifications happening over few minutes such as shown in~\cite{kagan2022vitro,masumori2015emergence}. We conclude that the hypothesis $\tau_r\ll \Delta t$ is satisfied. Furthermore, this estimate of $\Delta t$ would imply that running  hundreds of learning cycles would take few hours. We are well aware that these considerations are speculative and indicative of orders of magnitude at best.

\paragraph*{Intensities of stimulations.}
As emphasized above, having a large range of stimulation values at our disposal is crucial for reaching the target task at the end of learning. The values of $f_{max}$ we have employed in Tasks 1 \& 2 are a few tens of $r_{max}$, see Figs.~\ref{fig:fig4}(c) and \ref{fig:fig6}(d); details about how to convert input currents into rates can be found in SM, Section~\ref{supp:model1}. Ultra-fast optogenetics can elicit spikes in genetically-modified neurons with high probability (for sufficient light pulse intensity) up to hundreds of Hz~\cite{opto2}. Hence, ratios $f_{max}/r_{max}$ close to unity are experimentally accessible. It remains to be seen whether these techniques could be applied to organoids with similar results.

\paragraph*{Dependence on plasticity mechanism.} From a modelling point of view, we have resorted to a coarse-grained description of activity at the level of firing rates, rather than of spiking events. Both descriptions are compatible with one another, at the plasticity~\cite{izhikevich2003relating,bush2010reconciling,yger2015models} and activity~\cite{ostojic2011spiking,rueckauer2018conversion,sanzeni2020response} levels, to the point that artificial spiking networks may be trained by proxy rate-based models~\cite{kheradpisheh2022spiking}.   Plasticity at the spike level is a complex process involving multiple timescales, delayed effects and spike train patterns~\cite{froemke2002spike}, possible co-existing with non-synaptic plasticity~\cite{mozzachiodi2010more}, different effects depending on the origin of the neuron~\cite{caporale2008spike} and meta-plasticity~\cite{yger2015models}. The coarse-grained plasticity rules we consider respect some general principles: associativity (covariance rule with a threshold, compatible with experiments~\cite{lim2015inferring}), locality (dependence on the post and pre-synaptic neurons~\cite{caporale2008spike}), and homeostasis~\cite{zenke2017temporal}. In addition, we test the robustness of our approach with two different plasticity scenarios, based on Hebbian and anti-Hebbian rules for inhibitory synapses in, respectively, Tasks 2 and 3; see SM, Section~\ref{aht2} for reverse choices of the rules for inhibitory synapses. In this context, Hebbian and anti-Hebbian learning rules induce, respectively, the weakening and the strengthening of connections between inhibitory neurons with correlated activity, and result in increased or diminished~\cite{turrigiano2000hebb} activity. 
It would be interesting to exhaustively study the performances of our stimulation protocol over a wider class of plasticity rules. 

\paragraph*{Relevance for neuroscience.} As emphasized above, our modelling approach, besides being an oversimplification of the processes taking place in neurons and in their connections, focuses on {\em in vitro} circuits. Such circuits share many features, at the molecular and cellular levels, with their counterparts in the brain of animals, but also differ in fundamental ways. In particular, our learning protocol, based on control stimulations computed externally, was designed to meet bioengineering contexts and goals, but is {\em a priori} not directly relevant for neuroscience, e.g. to understand how animals learn in realistic situations.

\noindent
From this point of view, the interpretation of the training stimulation sequence is an important point to consider. We have observed that, in the case of the structural Task 2, the  structure of the training stimulation adapts to the ring symmetry of the target (Figs.˜\ref{fig:fig6}(d)\&(e)). Briefly speaking, the stimulation protocol engraves Fourier modes on the ring at finer and finer scales as (training) times passes on. This finding shows the deep connection between the spatio-temporal pattern of stimulations and the structure of the target network. Interestingly, the possibility of shaping a network with structured activity patterns is reminiscent of what is speculated to happen during postnatal development, for instance in the visual cortex~\cite{danka2020postnatal}. It would be interesting to further investigate this analogy,in particular if the activity driving neural development could be interpreted as the result of some sort of control optimization.

\section*{Methods}

\subsection*{Network structure and activation function}
The fractions of excitatory and inhibitory subpopulations are equal to, respectively, 80\% and 20\%. The synaptic structure is encoded in the binary matrix $C$: neuron $j$ is connected to neuron $i$ if $C_{ij}=1$, and is not if $C_{ij}=0$. 
The structure matrix $C$ is random and fixed during the training, while the intensities $J_{ij}$ of the existing connections ($C_{ij}=1$) vary over time according to Eq.~\eqref{eqpl}. 

The neuron activation $\Phi$ is a monotonously increasing function of the input  $x$ (which is dimensionally a frequency, see SM) and is chosen to be
\begin{multline}
    \Phi(x) = r_{max}\; \frac{ \psi(x)}{1+\psi(x)},
\qquad 
\text{where} \\ 
\psi(x)=\frac{r_0}{r_{max}}\, \log\left[1+\exp\left(\frac x{r_{0}}\right)\right]
\end{multline}
is a differentiable and invertible sigmoid function. Here, $r_{max}$ is the maximal firing rate reached for large inputs, and $r_0$ is the magnitude of the  spontaneous firing activity in the absence of input (SM, Section~\ref{supp:model}). 

\subsection*{Relaxation dynamics of the network activity}
While single-neuron activities are associated with the time constant $\tau_n$ in Eq.~\eqref{phi11}, the effective relaxation time $\tau_r$ can be longer due to network effects. We estimate $\tau_r$ by linearizing the dynamics of the rates around their stationary values, $r_i$. Denoting by $\mathbf{D}$ the $N$-dimensional diagonal matrix with elements $D_{ii}=\Phi'(r_i)$ and zero outside the diagonal, we have
\begin{multline}\label{defd}
    \tau_r = \frac{\tau_n}{\rho_{min}} \quad,\quad \text{where} \\ 
    \rho_{min} = \begin{aligned}
        &\text{smallest real part} \\ 
        &\text{of the eigenvalues of } \mathbf{Id} - \mathbf{D} \cdot \mathbf{J} 
    \end{aligned} \ .
\end{multline}
Estimates of $\tau_r$ are shown for the different tasks in SM Fig.~\ref{figS:fig1}

\subsection*{Inference of connectivity}

\paragraph**{Initialization: estimate of the synaptic structure $C$ and of the neuron types $\mathbf{\epsilon}$.} The support of the synaptic interactions, $C_{ij}=0,1$ and the nature of neurons, $\epsilon_i=E/I$, are initially unknown. To determine their values we start to probe the network activities $r_i^\nu$ in response to $n_{probe}$ randomly chosen stimulation patterns $f_{i}^\nu$, and rewrite the stationary relations in Eq.~\eqref{eq:stat} as
\begin{equation}\label{eq:cont2}
\sum_{j} J_{ij} \; r_j^\nu = t_i^\nu\ ,\quad \text{where} \quad t_i^\nu =\Phi^{-1}(r_i^\nu)-f_i^\nu 
\end{equation}
for all $i,\nu$. As the initial number of probing stimulations does not have to be small (plastic changes would simply affect the starting state of the connectivity for the training phase), we can choose  $n_{probe}\geq N$, and the linear system above generically determines the matrix $\mathbf{J}$ in a unique way. 
To avoid large numerical errors on currents through $\Phi^{-1}$, the probing stimulations $f_i^\nu$ are chosen to be large enough to avoid the presence of low firing rates ($r\simeq 0)$, see SM, Section~\ref{supp:model}.

Once $J$ has been estimated, we  estimate the identity of neurons and the support of the connectivity through, respectively, 
\begin{equation}
\epsilon_i=\left\{ \begin{array}{c c c} E &\mbox{ if }&\sum_j J_{ji} > 0 \\ I &&\mbox{ otherwise} \end{array}\right.
\end{equation}
and
\begin{equation}
C_{ij}=\left\{ \begin{array}{c c c}
1 &\mbox{ if } &|J_{ij}|>J_{min}\\ 0 &&\mbox{ otherwise} \end{array}\right.\ ,
\end{equation}
where $J_{min}$ is a small threshold, set to $10^{-6}$. These estimates of $\mathbf{\epsilon}$ and $\mathbf{C}$ are left unchanged throughout the training process.

\paragraph**{Updating the connectivity estimate after a training period.}
Let $\mathbf{J}_{i}$ be the vector of connections incoming onto neuron $i$ prior to a training stimulation period of duration $\Delta t$; the dimension of $\mathbf{J}_i$ is equal to $d_i=\sum _j C_{ij}$. After the training period the vector connections is $\mathbf{J}_i'$. To estimate this vector we probe the network with $n_{probes}$ stimuli, and record the corresponding activities. 
We call $\mathbf{R}$ the $(n_{probes}\times d_i)$-dimensional matrix of activities $r_j^\nu$ and $\mathbf{t}_i$ the $n_{probes}$-dimensional vector of components $t_i^\nu$ appearing in Eq.~\eqref{eq:cont2}. The solution to the constraints $\mathbf{R}\cdot\mathbf{J}'_i=\mathbf{t}_i$  is therefore not unique. To lift this ambiguity we select the solution $\mathbf{J}'_i$ closest (in terms of $L_2$ norm) to the previous estimate $\mathbf{J}_i$. Formally, we obtain \cite{montgomery2021introduction}
\begin{equation}\label{eqJupdate}
\mathbf{J}'_i=\boldsymbol{\Pi}^\perp_i \cdot \mathbf{J}_i + \mathbf{R}^+\cdot \mathbf{t}_i \ ,
\end{equation}
where $\mathbf{R}^+$ denotes the pseudo-inverse of $\mathbf{R}$ and $\boldsymbol{\Pi}^\perp_i =\text{Id}-\mathbf{R}^+\cdot\mathbf{R}$ is the projector orthogonal to the $n_{probes}$-dimensional space spanned by the rows of $\mathbf{R}$. This updating step is then iterated over the $N$ neurons $i$ to obtain all the vectors $\mathbf{J}'_i$.

\subsection*{Regularization of the connectivity matrix}

A common practice in machine learning is to introduce a regularization cost over the model parameters, hereafter denoted as $U_{reg}(\mathbf{J})$,  to smooth out the training trajectories. In addition, regularization can help ensure that the neural activities reach their stationary values quickly (under fixed stimulation) through the network dynamics in Eq.~\eqref{phi11}, which is important for training and for connectivity estimation.

We impose regularization to bound the modulus $\sigma(\mathbf{J})$ of the largest singular value of the connectivity matrix.
The rationale is that a sufficient condition for the network dynamics to have a unique activity stationary state and converge exponentially fast to this unique state is that $\sigma(\mathbf{J})<1$, see Eq.~\eqref{defd} and SI, Section 2.2. For each singular value $\lambda_k$ of $\mathbf{J}$, the regularization cost is approximately zero if $\lambda_k\ll 1$, and grows linearly as $g_2\,\lambda_k$ for $\lambda_k\gg1$. We implement this dependence through a softplus function:
\begin{equation}\label{ureg}
U_{reg}(\mathbf{J}) =\frac{g_2}{g_1}\,\sum_{k=1}^{N^2} \log\left(1+e^{\,g_1\,(\lambda_k-1)}\right) 
\end{equation}
with $g_2=2, \ g_1=10$. This regularization cost can be explicitly differentiated with respect to the entries of the connectivity matrix, see below.

\subsection*{Determination of the optimal stimulation $\mathbf{f}^*$}

\paragraph**{Derivative of the costs with respect to connections and time.} The task cost $U_{task}$ depends on the connectivity $\mathbf{J}$, both directly and indirectly through the stationary activity $\mathbf{r}$. The expression for the total derivative of $U_{task}$ with respect to an entry of $\mathbf{J}$ is, according to chain rule, 
\begin{equation}\label{dU1}
    \frac{d U_{task}}{d J_{ij}}= \frac{\partial U_{task}}{\partial J_{ij}} + \sum_{k} \frac{\partial U_{task}}{\partial r_k} \frac{\partial r_k}{\partial J_{ij}} 
\end{equation}
which requires knowledge of the derivatives of the neuron firing rates. These can be computed using the implicit function theorem~\cite{rudin1953principles} applied to the stationary equations for the dynamics, with the result:
\begin{equation}\label{dU2}
    \frac{\partial r_k}{\partial J_{ij}} = \big[\big(\mathbf{Id} -\mathbf{D}\cdot\mathbf{J}\big)^{-1}\big]_{k,i}\, \mathbf{D}_{ii}\,  r_j \ ,
\end{equation}
where $\mathbf{D}$ was first used in Eq.~\eqref{defd}. The gradient of the regularization cost, $U_{reg}$ in Eq.~\eqref{ureg}, can be easily expressed using  the derivatives of the singular values:
$\frac{d\lambda_k}{dJ_{ij}}=u_{k,i}\, v_{k,j} $, where $\mathbf{u}_k$ and $\mathbf{v}_k$ denote, respectively, the left and right eigenvectors of $\mathbf{J}$ associated to $\lambda_k$.

In addition, the gradient component associated to the connection $J_{ij}$ is set to zero if $C_{ij}=0$ (no connection is actually present), or if $J_{ij}=0$ and descending the gradient would violate the constraint on the sign of the interaction resulting from the pre-synaptic neuron type, {\em i.e.} $\frac{dU}{dJ_{ij}}>0$ if $\epsilon_j=E$ or $\frac{dU}{dJ_{ij}}<0$ if $\epsilon_j=I$. Notice that knowledge of the gradients of $U_{task}$ and $U_{reg}$ with respect to $J_{ij}$ gives access to the time derivatives of these costs (under fixed stimulation) through the generic formula
\begin{equation}\label{dU3}
    \dt{U} = \sum_{i,j} \frac{\partial U}{\partial J_{ij}}\;   {\dt J}_{ij} \ ,
\end{equation}
where the latter term can be directly obtained from the dynamics over synapses described by Eq.~\eqref{eqpl}.

\paragraph**{Objective function and gradient over stimulations.}
The total cost $U$ is defined as the sum of the task and regularization costs, $U=U_{task}+U_{reg}$. To determine the best control stimulation $\mathbf{f}$ over the next period we search for the minimum of the modified change 
\begin{equation}
\Delta U(\mathbf{f};\gamma)=U\big(\mathbf{J}+\Delta \mathbf{J}(\mathbf{f}) \big) -U\big(\mathbf{J}\big)+\Delta t\; \gamma\;\|\Delta \mathbf{J}(\mathbf{f})\|^2 \label{deltaU}
\end{equation}
where $\mathbf{J}$ is the matrix of connectivity estimated through probing at the end of the previous control stimulation period, $\gamma=1/((\bar J/10)^2\,c\,N^2)$ and
\begin{equation}
    \Delta \mathbf{J}(\mathbf{f})=[\mathbf{J}+\Delta t\,\dot{\mathbf{J}}(\mathbf{f})]_{\boldsymbol{\epsilon}}-\mathbf{J}
\end{equation}
where $[\cdot]_{\boldsymbol{\epsilon}}$ indicates that connections have been clipped to respect their excitatory/inhibitory types $\epsilon$ imposed by the pre-synaptic neurons. The last term in Eq.~\eqref{deltaU} constrains the change in the interactions to be small over the period of stimulation, effectively introducing a saturating non-linearity over the gradients.

Computing the gradient of $\Delta U$ in Eq.~\eqref{deltaU} with respect to the control stimulation components $f_i$ requires the expressions for the derivatives of the costs with respect to the connections in Eqs.~\eqref{dU1},\eqref{dU2}, and with respect to time, see Eq.~\eqref{dU3}. In  addition, the derivatives of the firing activities are needed. Using again the implicit function theorem we obtain, see Eq.~\eqref{dU2},
\begin{equation}
    \frac{\partial r_k}{\partial f_i}=\big[\big( \mathbf{Id}-\mathbf{D}\cdot\mathbf{J}\big)^{-1}\big]_{k,i}\, \mathbf{D}_{ii}\ .
\end{equation}
To enforce the conditions $f_{min}<f_i<f_{max}$ we set the gradient components of $\Delta U_{tot}$ to zero when these boundaries are met. 

\paragraph**{Minimization of the cost.}
The optimal control $\mathbf{f}^*$  can be found through an iterative descent of $\Delta U$ in Eq.~\eqref{deltaU}. The control is updated along minus the gradient of $\Delta U(\mathbf{f},\gamma)$, and the resulting $f_i$'s are clipped if the boundaries $f_{min}$ or $f_{max}$ are crossed. The iterative process halts if the decrease of $\Delta U(\mathbf{f},\gamma)$ is smaller (in absolute value) than some threshold, or the expected decrease (based on a fit of the previous steps in the process) is too low. Then, if $\Delta U(\mathbf{f},0)$ is negative (i.e. the cost function is decreasing), the solution is accepted, while the gradient descent is repeated otherwise.

Details about the implementation, with the pseudocode for the optimization loop and the values of the hyper-parameters involved  can be found in SM, section~\ref{supp:optim}.

\subsection*{Task 1}

\paragraph**{Initialization and setting.}
Elements of the structure matrix $C_{ij}$ are randomly set to 0 or 1 with respective probabilities $1-c_E$ and $c_E$ for excitatory columns ($J$ such that $\epsilon_j=E$) and $1-c_I$ and $c_I$ for inhibitory ones ($\epsilon_j=I$). 
To ensure that the task is not easily implementable no direct connection between the $N_{in}$ input neurons and the $N_{out}$ output neurons is present.
Initial connections $J_{ij}$ are drawn uniformly at random in the ranges $[0, J_0]$ if $\epsilon_j>0$ or $[-J_0,0]$ if $\epsilon_j<0$. See SM, Section~\ref{supp:model} for parameter values.

\paragraph**{Task-associated cost.}
We define a set of $n_{task}$ input stimulations $\mathbf{f}^\mu$ (defined over the $N_i$ input neurons) and output binary activation $\pmb{\sigma}^\mu$ (over the $N_o$ output neurons), with $\mu=1,...,n_{task}$. The goal of training is that output neuron $i$ should display High or Low firing activity in response to input $\mu$ when, respectively, $\sigma^\mu _i=H$ or $L$. In practice we enforce that the firing rates associated to neurons with low ($\sigma=L$)
and high ($\sigma=H$) activities should differ by more than a prescribed gap $\delta r= 0.12\, r_{max}$. The task cost reads
\begin{widetext}
\begin{equation}\label{maincosttask}
    U_{task}(\mathbf{J})=
\frac{\displaystyle{\sum_{(\mu,i):\sigma^\mu _i=L}\ \ \sum_{(\nu,j):\sigma^\nu_j=H} \ \Delta(\mu,i;\nu,j) \; e^{\gamma\,\Delta(\mu,i;\nu,j)} }} {\displaystyle{\sum_{(\mu,i):\sigma^\mu _i=L}\ \ \sum_{(\nu,j):\sigma^\nu _j=H}\  e^{\gamma\,\Delta(\mu,i;\nu,j)}}}
\end{equation}
\end{widetext}
where $\gamma=1/(\delta r/2)^2$ and we have introduced the mismatch  
\begin{equation}\label{costclass}
\Delta(\mu,i;\nu,j)=h_2 \big(r_i^\mu-r_j^\nu +\delta r \big) 
\end{equation}
with
\begin{equation}
h_2(u) = u^2 \ \text{if} \ u>0, \ 0 \ \text{otherwise}.
\end{equation}
The exponential factors  give more weights to large mismatches $\Delta$ in the expression of the cost. The dependence of the cost on $\mathbf{J}$ (and on the input stimulations $\mathbf{f}^\mu$) in Eq.~\eqref{costclass} is implicit through the firing rates $r_i$. Minimizing $U_{task}$ is thus equivalent to separating the activities of the $H$ and $L$ neurons as required by the classification task, {\em i.e.} making all $\Delta(\mu,i;\nu,j)= 0$.

\paragraph**{Analysis of input and output spaces.} 

To characterize how the association mechanism emerges across training, we convert digits $\sigma_{i}^\mu=$High/Low to 1/0 valued arrays. These digits span a $n_{pairs}=4$-dimensional space. Given a stimulation $\mathbf{f}$, we call $\mathbf{r}_{out}(\mathbf{f})$ the output activity and 
\begin{equation}
\|\mathbf{r}_{out} ^\perp(\mathbf{f})\|^2=\frac{ \min_{\boldsymbol{\beta}}\left\|\sum_{\mu=1}^{n_{task}}\boldsymbol{\sigma}_\mu^i\beta_\mu-\mathbf{r}_{out}(\mathbf{f})\right\|^2}{ {\|\mathbf{r}_{out}(\mathbf{f})\|^2}}\ .
\end{equation}
the normalized squared length of its projection orthogonal to the digit space. In Fig. \ref{fig:fig5}(a), we report this quantity for: (1) generic inputs  $\mathbf{f}$ with 0,1 entries with 50-50 probability, and (2) random, uniform convex combinations of the inputs associated to the digits:  $\mathbf{f}=\sum_{\mu=1}^{n_{task}}\alpha_\mu \mathbf{f}_\mu$
with $\alpha_\mu\geq0$ and $\sum_\mu\alpha_\mu=1$. 

\subsection*{Task 2}

\paragraph**{Initialization and setting.} The connectivity matrix is initialized as in the classification task above, but no {\em a priori} restriction is imposed to the structure, {\em i.e.} $c_E=c_I=1$. See SM, Section~\ref{supp:model} for parameter values.

\paragraph**{Task and cost.} The structural cost is defined in Eq.~\eqref{cost_struct}. Weights are defined as $w_{\epsilon_i,\epsilon_j}=\big(\sum_{ab}[J^{target}_{\epsilon_i\epsilon_j}]_{ab}^2\big)^{-1}$, 
where $J^{target}_{\epsilon_i\epsilon_j}$ is the sub-matrix of $\mathbf{J}^{target}$ connecting neurons of type $\epsilon_j=E/I$ to neurons of type $\epsilon_i=E/I$. This re-weighting ensures that large, {\em e.g. 
} $E\to E$, and small, {\em e.g. 
} $I\to I$ blocks equally contribute to the cost, and are simultaneously and properly learned during the training process. 

The target connectivity is defined as follows. Neurons are associated angles on two rings, depending on their types $\epsilon=E$ or $I$. The angle attached to the excitatory neuron $i$ is $\theta^E_i=2\,\pi\,\frac{i}{N_E}$, where $N_E$ is the number of excitatory neurons. A similar formula holds for the $N_I$ inhibitory neurons: $\theta^I_i=2\,\pi\,\frac{i}{N_I}$. The target synaptic connection from neuron $j$ to $i$ reads
\begin{equation}\label{kk1}
    J^{target}_{ij}=J_{\epsilon_i,\epsilon_j} \times \mathcal{M}\big( \theta_i^{\epsilon_i}-\theta_j^{\epsilon_j}-\pi\, \delta_{\epsilon_i,I}; K_{\epsilon_i,\epsilon_j}\big) \ ,
\end{equation}
where $\delta_{.,.}$ denotes the Kronecker delta. The angular modulation is done through von Mises function, 
\begin{equation}\label{kk2}
    \mathcal{M}(\Delta \theta;K)=e^{- K + K\,  \cos(\Delta\theta)} \ ,
\end{equation}
which is maximal for vanishing angular separation ($\Delta \theta=0$) and decays over the width $\simeq K^{-1/2}$. Parameter values are listed in  SM, Section~\ref{supp:model}.

\paragraph*{Receptive fields.} We apply a weak local input stimulation  $f_i=0.06\,r_{max}$, $f_{i\pm1}=0.028\,r_{max}$, $f_{i\pm2}=0.012\,r_{max}$, $f_{i\pm3}=0.004\,r_{max}$ to induce polarization of neuron $i$, and repeat the process for all $i$.
Results for the activities are averaged over 100 trials.

\bibliographystyle{apsrev4-2}
\bibliography{learning_references}

\newpage


\newpage

\section*{Acknowledgements }

We thank S. Wolf for many useful discussions, and D. Chatenay for reading the manuscript. This research was supported by the European Union’s Horizon 2020 research and innovation program under Grant No. 964977 (project NEU-ChiP).

\section*{Declaration of interests }
The authors declare no competing interests.


\section*{Author contributions statement}

All authors planned and conceived the study. F.B. designed the algorithms, ran the simulations, and analyzed the results. F.B. and R.M. wrote the manuscript;  All authors reviewed the manuscript. 

\FloatBarrier
\onecolumngrid

\newpage
\thispagestyle{empty} 
\mbox{} 
\newpage 
\appendix

\begin{center}
    \rule{\textwidth}{0.5mm} \\[1ex] 
    {\LARGE \textbf{Supplementary Material}} \\[1ex]
    \rule{\textwidth}{0.5mm} 
\end{center}

\section{Model parameters: values and biological significance}\label{supp:model}

\subsection{Values}

\paragraph{Task 1.} We use $c_E=0.2$ and $c_I=0.5$. 
The parameters are $J_0=0.015$, $\theta(E)=0.08\,r_{max}$, $\theta(I)=0.12\,r_{max}$, $\eta(E)=1$, $\eta(I)=-1.2$, and $\beta_1=0.8$, $\beta_2=0.6\,(r_{max}/\bar J)^2$, $\theta_0(E)=\theta_0(I)=0.8\, r_{max}$ and $\bar J=0.25$, $f_{max}=-f_{min}=0.2\,r_{max}$. Note that, in this setting, the associative behaviour of excitatory synapses is triggered for lower values of activity ($\theta(E)<\theta(I)$); however, inhibitory potentiation is favored by $|\eta(I)|>\eta(E)$. For the sake of inferring $J$, we use $n_{probe}=10$ plus the four input stimulations related to the task. The control stimulation lasts $\Delta t=0.003\,\tau_s$. Probing for estimating connectivty is done by choosing $f_i=0,f_{max}$ with equal probabilities.

\paragraph{Task 2.} Parameters are chosen as follows: $\theta(E)=0.2\,r_{max}$, $\theta(I)=0.16\,r_{max}$, $\eta(E)=1$, $\eta(I)=1$, and $\beta_1=0.8$, $\beta_2=1.25\,(r_{max}/\bar J)^2$, $\theta_0(E)=\theta_0(I)=0.16\, r_{max}$ and $\bar J=0.1$, $f_{max}=-f_{min}=0.4\,r_{max}$. We use $n_{probe}=10$ and $\Delta t= 0.003\,\tau_s$. Probing for estimating connectivity is done with $f_i$ distributed uniformly in $[f_{max}/2,f_{max}]$ for excitatory neurons and $[0,f_{max}/2]$ for inhibitory neurons; further inforamtion is available in SI, Section~\ref{supp:stimtask2}.

The interaction strengths $J$ and modulations $K$ entering the interaction kernels, see Eqs.~\eqref{kk1}\eqref{kk2} in the main text, are chosen to be
$J_{E,E}=J_{I,E+}=-J_{E,I}=-J_{I,I}=0.1$, 
and
$K_{E,E}^{-1/2}=0.08$, $K_{I,E}^{-1/2}=0.05$, $K_{E,I}^{-1/2}=0.15$, $K_{I,I}^{-1/2}=0.1$. These parameter values are compatible with the existence of two self-sustained bumps of activity, localized and diametrically opposed on the two rings.

In the cost $U$ in Eq.~\eqref{cost_struct}, the weight factors are chosen as $w_{\epsilon_i\epsilon_j}=\big(4\sum_{ab|\epsilon_a=\epsilon_i,\epsilon_b=\epsilon_j}[J^{target}_{ab}]^2\big)^{-1}$ to ensure that the cost assesses the average relative errors over the four classes of connections $E/I\to E/I$.

\subsection{Current-to-firing rate activation function}\label{supp:model1}
We discuss here the expression for the activation function $\Phi$ used in Eqs.~(1)~and~(8) in the main text.
Three regimes must be distinguished.

\paragraph{From input currents to firing rates.}
Consider the linear integrate-and fire model of a spiking neuron. In the absence of leakage (zero conductance), the membrane potential $V$ obeys the following dynamical equation
\begin{equation}
    C\,\dot V=I  ,
\end{equation}
where $I$ is the input current (considered to be constant over time) and $C$ the capacitance. 
The neuron fires with rate
\begin{equation}\label{itor}
    r(I)=MF\times I \quad \text{with}\quad MF= \frac 1{C(V_{threshold}-V_{rest})}\ ,
\end{equation}
and $V_{rest}$ and $V_{threshold}$ are, respectively, the rest and instability potentials. Equation~\eqref{itor} shows that input currents can be expressed in units of firing rates through the rescaling by the multiplicative factor $MF$: $I\to \tilde I = MF\times I$. This is the convention we adopt throughout our paper.

An order of magnitude for $MF$ is given by the slope of the current-to-frequency relationship for human pyramidal cells~\cite{moradi2021diversity}, with the result $MF\approx 50$ Hz/nA. Compatible values are reported for pyramidal cells in the primary visual cortex of cats~\cite{nowak}, with $MF\approx100-500$ Hz/nA.

\paragraph{High activity regime.}
The maximal firing rate $r_{max}$ is the inverse of the refractory period, with values of the order of a few hundreds Hz. The linear current-rate relation in Eq.~\eqref{itor} can be simply modified as
\begin{equation}\label{ilarge}
    r(\tilde I)=\frac{r_{max}\;\tilde I}{r_{max}+\tilde I} \qquad \text{(\ large $\tilde I$\ )}\ .
\end{equation}
This expression fixes both the maximum firing rate and the way $r$ reaches it for large currents $\tilde I$. Many effects such as fatigue, temporal fluctuations (in the concentration of neurotransmitters),  or short term plasticity are not taken into account.

\paragraph{Low activity regime.}
For low external input $\tilde I$, noisy synaptic effects can become predominant and set a baseline firing activity $r_{low}$ for the neuron. 
We can interpolate between the linear response in Eq.~\eqref{itor} and this low current, baseline activity regime through the introduction of a smooth function,
\begin{equation}\label{ilow}
    r(\tilde I)= r_0 \;\log\big(1+\exp(\tilde I/ r_0)\big)\qquad \text{(\ small $\tilde I$\ )}\ .
\end{equation}
where $r_0=r_{low}/\log 2$. 
Note that $r_0$ also defines the scale at which  $\tilde I$ can be considered small. We expect $r_0$ to range between 0.1 and 1Hz~\cite{moradi2021diversity}. The ratio between the maximal and baseline firing rates is thus of the order of $r_{max}/r_0=50-2000$. In this paper, we use $r_{max}/r_0=250-500$.

\paragraph{Complete activation function.} The activation function described in Eq.~(8) of Methods, main text, interpolates between the low, intermediate, and high activity regimes corresponding, respectively, to Eqs.~\eqref{ilow},~\eqref{itor},~\eqref{ilarge}. 

\subsection{Synaptic connections}

A pre-synaptic neuron, depending on its type, can give rise to small excitatory or inhibitory post-synaptic potentials, referred to as EPSP or IPSP, each time it emits a spike. The mean value of the resulting current incoming onto the post-synaptic neuron is, after rescaling by $MF$, equal to
\begin{equation}
    \tilde I = \frac{\delta V}{V_{threshold}-V_{reset}}\; r_{pre} \ ,
\end{equation}
where $\delta V$  is the value of the EPSP or IPSP, and $r_{pre}$ is the firing rate of the pre-synaptic neuron. We deduce the dimensionless expression of the synaptic connection, compare with Eq.~(1) in the main text,
\begin{equation}
    J = \frac{\delta V}{V_{threshold}-V_{reset}} \ . 
\end{equation}
Typical values for $|\delta V|$ range from tens to hundreds $\mu$V~\cite{epsp}. As $V_{threshold}-V_{rest}$ is about $20$~mV, we find $J$ to be of the order of $5.10^{-4}$ to $5.10^{-2}$. These values are consistently lower than the soft bounds $\bar J=0.25$ in Task~1 and $\bar J=0.1$ in Task~2.

\subsection{Control stimulation}

\paragraph{Amplitude.} The second source of input in our model is the external control $f$, see Eq.~(1) in the main text. Literature reports accurate control of the activity of genetically-modified neurons through light activation, with firing rates up to 30 Hz with Channelrhodopsin-2~\cite{opto1} and up to 100 Hz with faster Chrimson variants~\cite{opto2}. The firing rates can be controlled both through the frequency of light pulses and the illumination power. These values are compatible with ratios $f_{max}/r_{max}$ reaching a few tenths, as considered in this work.
Inhibitory stimulations can also be obtained with optogenetic tools, resulting in strong hyperpolarization of cells by more than 10~mV, and strong reduction (50\%) in the firing activity~\cite{opto3}.

\paragraph{Duration.} Due to the choice of the parameters and the values of firing rates, all terms on the right hand side of Eq.~(\ref{eqpl}) in the main text take values of the order of unity. The order of magnitude of the time required for a variation $\delta J$ in a synaptic connection is therefore
\begin{equation}
    \delta t \simeq \tau_s\; \delta J\ .
\end{equation}
Substituting $\delta J$ with $\bar J/10$ in the above equation, where $\bar J$ is the soft bound on the (absolute) value of connections, we find that a significant change to connection requires time $\delta t\simeq \tau_s\times \bar J/10\simeq 0.01\;\tau_s$. 

Hebbian plasticity is associated to a variety of times scales $\delta t$, ranging from few seconds to few minutes, with homeostatic effects taking place over a large interval of time scales too~\cite{zenke2017temporal}. Choosing $\delta t$ of the order of 1 minute gives $\tau_s$ in Eq.~\eqref{eqpl} of the order of a few hours. The duration of one cycle of control stimulation is then approximately equal to $\Delta t=0.003\; \tau_s\simeq 30$ sec. This time scale is consistently larger than the time $\tau_r$ for convergence of the network activity, see next Section and Discussion in the main text.

\newpage

\section{Stationary activity of the rate model}~\label{supp:stationarity}

\subsection{Convergence of dynamics}

The stationary states of activity are the roots of the $N$ coupled implicit equations
\begin{equation}\label{steqsupp2}
\mathbf{r}=\Phi\big(\mathbf{J}\cdot\mathbf{r}+\mathbf{f}\big)\ .
\end{equation}
Solutions, in general, are not unique. Moreover, since the external input $\mathbf{f}$ is arbitrary, uniqueness may hold for some $\mathbf{f}$ and not for others. However, if  the largest singular value of $\mathbf{J}$, $\sigma(\mathbf{J})$, is smaller than unity, the solution $\mathbf{r}^\star$ of~\eqref{steqsupp2} is guaranteed to be unique\footnote{$\sigma(\mathbf{J})$ coincides with the spectral radius of $\sqrt{\mathbf{J}^T \mathbf{J}}$}. Moreover, $\mathbf{r}^\star$ is a globally attractive fixed point of the dynamical equations (expressing time in units of $\tau_n$ to lighten notations):
\begin{equation}\label{relaxeqsupp2}
\dot {\mathbf{r}}=-\mathbf{r}+\Phi(\mathbf{J}\cdot\mathbf{r}+\mathbf{f}).
\end{equation}
and, therefore, no limit cycle can exist. 

\paragraph{Uniqueness of solution.} Consider the discrete-time dynamics
\begin{equation}\label{timedisceq}
\mathbf{r}(k+1)=\Phi(\mathbf{J}\cdot\mathbf{r}(k)+\mathbf{f})\ .
\end{equation}
As $\sigma(\mathbf{J})=\max_{\mathbf{x}} \|J\,\mathbf{x}\|/\|\mathbf{x}\|<1$, the map $L_{J,f}:r\mapsto \mathbf{J}\cdot\mathbf{r}+\mathbf{f}$ is a global contraction for any $\mathbf{f}$. Moreover, since $|\Phi'|<1$, $\Phi$ is a global contraction, too. Hence, the map in~\eqref{timedisceq}, which can be written as $\Phi\circ L_{J,\mathbf{f}}$, is a global contraction. According to the contraction mapping theorem, Eq.~\eqref{timedisceq} has a unique fixed point given by $\mathbf{r}^\star=\lim_{k\to\infty}\mathbf{r}(k)$ for any $\mathbf{r}(0)$. This result gives us a practical and fast way to compute $\mathbf{r}^\star$.

\paragraph{Attractivity of solution. } We now show that the unique solution of the stationary solution is an attractive fixed point of the continuous-time dynamics. Assume $\mathbf{r}\neq \mathbf{r}^\star$ and note that, under dynamics~\eqref{relaxeqsupp2},
\begin{equation}
\frac{1}{2}\frac{d}{dt} \|\mathbf{r}-\pmb{\Phi}\|^2=(\mathbf{r}-\pmb{\Phi})^T\cdot \mathbf{M}(\mathbf{r})\cdot(\mathbf{r}-\pmb{\Phi})-\|\mathbf{r}-\pmb{\Phi}\|^2
\end{equation}
where we have used the shorthand notation $\pmb{\Phi}=\Phi(\mathbf{J}\cdot\mathbf{r}+\mathbf{f})$ and $M_{ij}(\mathbf{r})=\Phi'_i(\mathbf{J}\cdot\mathbf{r}+\mathbf{f})\, J_{ij}$. Note that since $\Phi'_i<1$ for all $i$, the element-wise multiplication by $\Phi'_i$ is a contraction. Then, if $\sigma(\mathbf{J})<1$, $\|\mathbf{J}(\mathbf{r})\cdot(\mathbf{r}-\pmb{\Phi})\|<\|\mathbf{r}-\pmb{\Phi}\|$, from which it follows that
\begin{equation}
\frac{d}{dt} \|\mathbf{r}-\pmb{\Phi}\|^2<0 \;\;\forall \mathbf{r}\neq \mathbf{r}^\star.
\end{equation}
We conclude that $L(\mathbf{r})=\|\mathbf{r}-\pmb{\Phi}\|^2$ is a global Lyapunov function for~\eqref{relaxeqsupp2} for any $\mathbf{f}$ in the sense that it satisfies Lyapunov's second method for stability; note that $L(\mathbf{r}^\star)=0$.   Hence, whatever the initial condition, $\mathbf{r}$ converges asymptotically to $\mathbf{r}^\star$ under the dynamics~\eqref{relaxeqsupp2}.

In practice, to find the fixed point, we  iterate~\eqref{timedisceq}. Note that $\sigma(\mathbf{J})<1$ is a sufficient but not a necessary condition. Violating it does not always appear to be too detrimental to convergence. While we softly enforce this condition in the classification Task 1, we violate it on purpose in the case of Task 2, since a continuous attractor is expected to have multiple solutions by construction.

\subsection{Relaxation time}

\begin{figure}[h!]
    \centering
     \includegraphics[width=1.\textwidth]{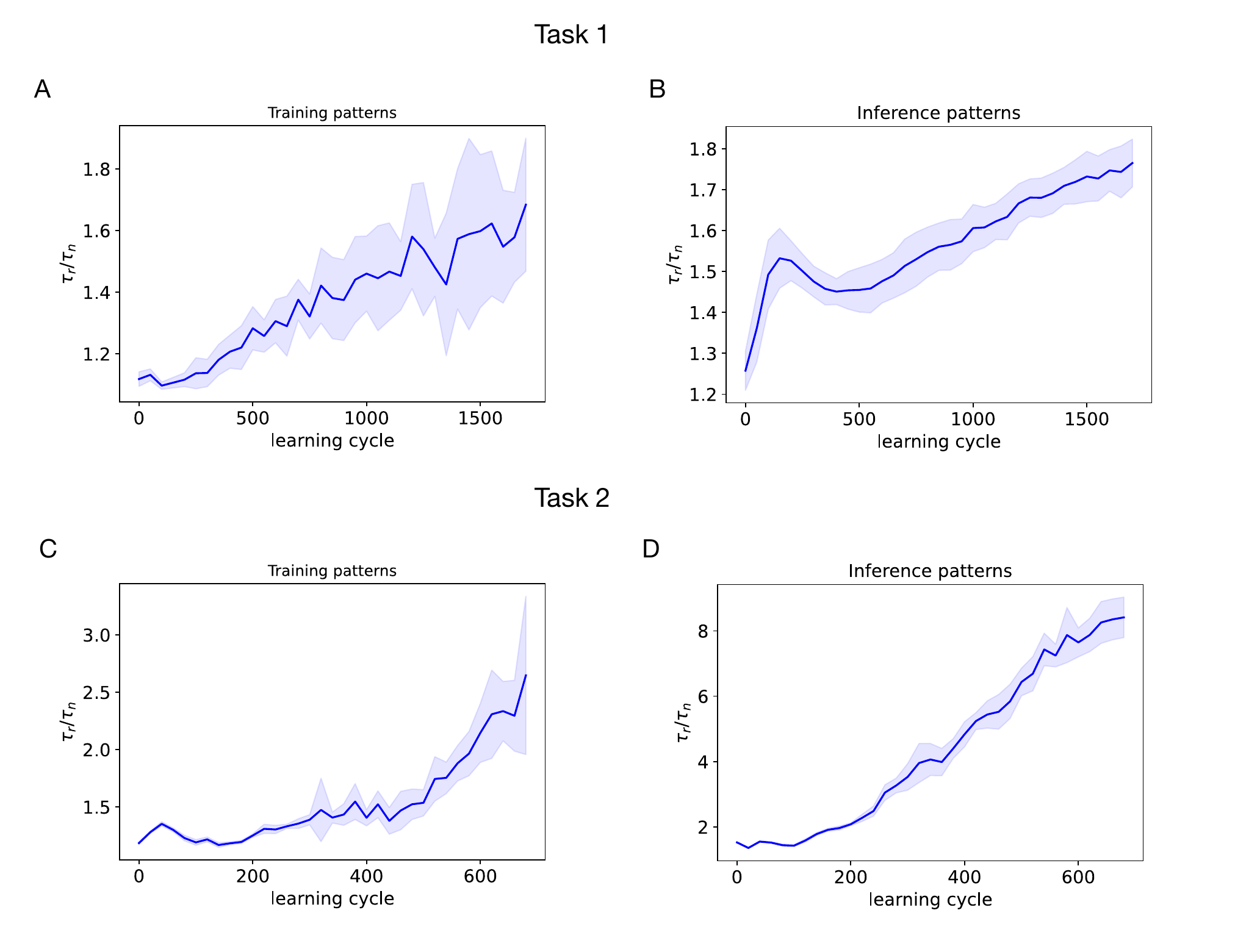}
    \caption {Relaxation times $\tau_r/\tau_n$ for Task 1 (top row, A and B),  and Task 2 (bottom row, C and D). In the left column we show results for the training patterns, while the outcome for random probing patterns (with random entries 0 or 1) used for connectivity estimation are shown in the right column. Results are averaged over 10 instances for training examples, and over 20 examples in the case of the inference patterns; the  blue shaded area corresponds to one standard deviation from the mean.  }
    \label{figS:fig1}
\end{figure}

We can also estimate how fast the firing rates converge towards the stationary solution. Linearizing the dynamics in Eq.~\eqref{relaxeqsupp2} to the first order in $\delta\mathbf{r}=\mathbf{r}-\mathbf{r}^\star$, we obtain  
\begin{equation}\label{linrelaxeqsupp2}
\frac{d}{dt} \delta\mathbf{r}=(-\mathbf{Id}+\mathbf{M}(\mathbf{r}^\star))\cdot\delta\mathbf{r}\ .
\end{equation}
Hence, the relaxation time $\tau_r$  is given by
\begin{equation}\label{linRT}
\tau_r=\frac{\tau_n}{1-\max\{\text{Real part of eigenvalues of }\mathbf{M}(\mathbf{r}^\star)\}}\ .
\end{equation}
The relaxation time $\tau_r$ includes network effects, and is expected to be larger than the single-neuron time scale $\tau_n$. Note that $\tau_r$ does not depend only on the connectivity $\mathbf{J}$, but also on the input $\mathbf{f}$. 
We show how the ratio $\tau_r/\tau_n$ varies during training in Fig.~\ref{figS:fig1}, both for Tasks 1 and 2. We observe that the introduction of a regularization over the connections for Task 1 consistently reduces the relaxation time. However, even in the absence of this extra cost as in Task 2, the value of $\tau_r/\tau_n$ remains moderate up to the end of the training phase.

\section{Optimization loop over the control}\label{supp:optim}
The control $\mathbf{f}$ can be found with a training loop that minimizes the auxiliary cost 
\begin{equation}W(\mathbf{f})=\frac{\Delta U(\mathbf{f})}{\Delta t}+\gamma   \|\Delta\mathbf{J}(\mathbf{f})\|^2
\end{equation} 
with
\begin{equation}\Delta U(\mathbf{f})=U(\mathbf{J}+\Delta \mathbf{J}(\mathbf{f}))-U(\mathbf{J})
\end{equation} 
and $\gamma=1/(\bar J/10)^2/cN$. Exit from the minimization loop takes place if $W$ is sufficiently low, {\em i.e.} below a hyper-parameter $W^{threshold}$, which is negative at the beginning to make sure that the algorithm does not get stuck immediately, but is then set to 0. This condition is checked once $W$ is not changing anymore, {\em i.e.} either the gradient of $W$ is zero or the running average of the differences $\Delta W$ of  $W$ between two successive loop iterations is small than a threshold $\Delta W^{threshold}$ (defining another hyper-parameter). If either of these conditions is verified, the loop is broken if $\Delta U<\Delta U^{threshold}$ and re-initialized with a new initial condition for $\boldsymbol{f}$ otherwise. A pseudocode for the training loop to find $\mathbf{f}$ at a given time step is provided below. A number of hyper-parameters and checks are involved in this optimization, which depend on the setting and can be found in the code.

\newpage
\vskip .3cm
\hrule
\vskip .3cm
\noindent
{\large Initialization:}

Take $\mathbf{f}_{old}$ from previous control stimulation (if there is one) and a random new one $\mathbf{f}_{random}$. Check which one has the lowest scalar product with $\partial\Delta U/\partial \mathbf{f}$ and pick this as the initial condition for the loop. This step aims at preventing the control to get stuck in local minima.

\vskip .3cm
\noindent
{\large Loop:}
\begin{enumerate}
\item compute $\displaystyle{\frac{dW}{d\mathbf{f}}}$.
\item If $\displaystyle{\frac{dW}{d\mathbf{f}}=0}$ or $\displaystyle{\langle\Delta W \rangle<\Delta W^{threshold}}$ then
\begin{itemize}
    \item if $\Delta U<\Delta U_{threshold}$ then exit
    \item else  reset. Randomly draw  a new $\mathbf{f}$  and reset the running average $\langle\Delta W \rangle$ to a high value.
\end{itemize}    
\item else update the control as follows
\begin{equation}
\mathbf{f}\leftarrow \mathbf{f}-\alpha \frac{dW}{d\mathbf{f}}
\end{equation}
where  the learning rate $\alpha$ is chosen to minimize $\Delta W$ by scanning multiple values; this step is necessary because the optimum learning rate  significantly changes with the connectivity $\textbf{J}$. Clip element-wise $\mathbf{f}$ in the range $[f_{min}-q,f_{max}+q]$ for a small value of $q$ (introduced to improve the numerical procedure).
\item compute $W$ and $\Delta U$ and update the running average $\langle\Delta W \rangle$.
\end{enumerate}
\hrule
\vskip .3cm

\begin{figure}[h!]
    \centering
     \includegraphics[width=.9\textwidth]{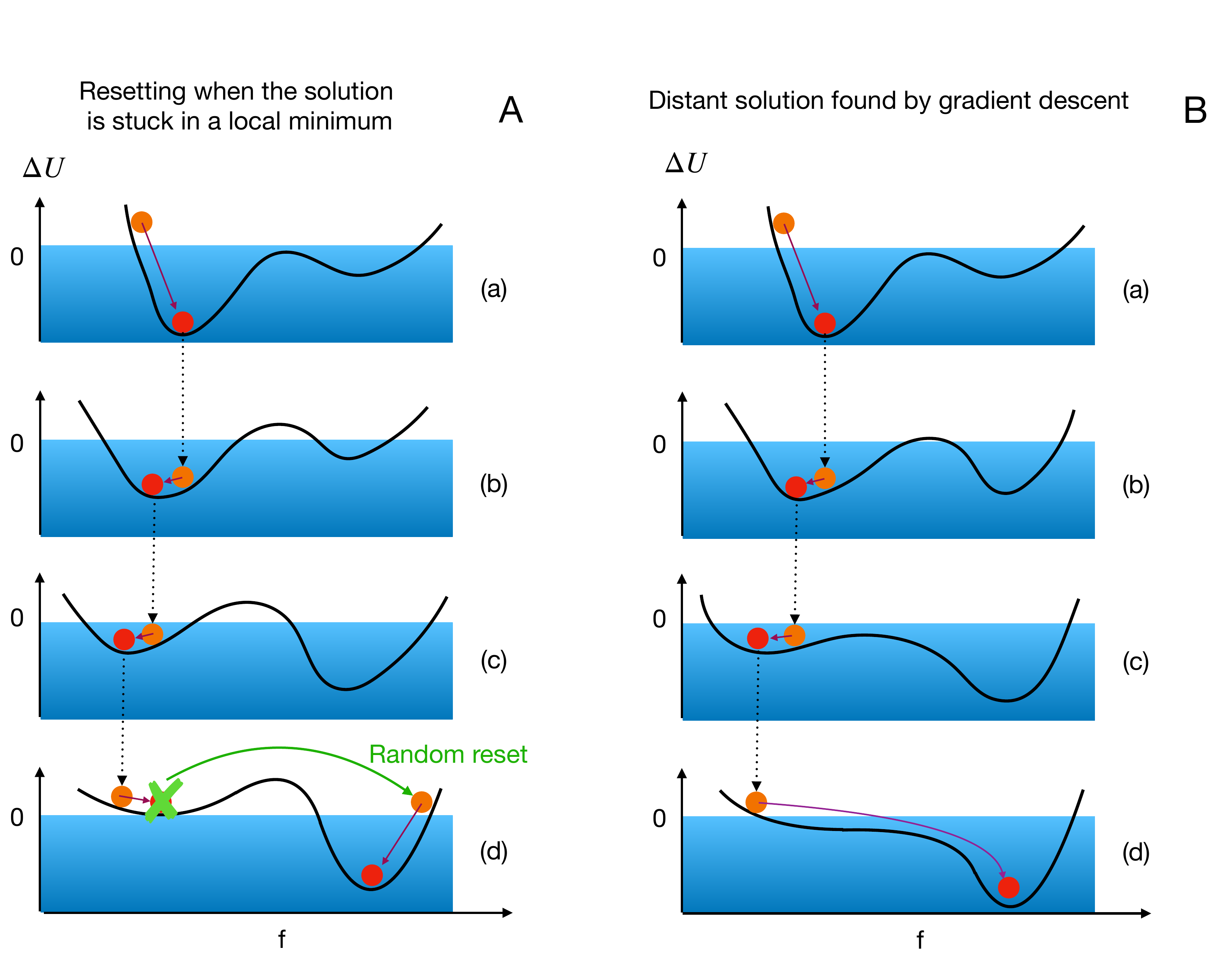}
    \caption {The two mechanisms generating large discontinuities in the control variable. Panel A: external reset. Panel B distant solution found by gradient descent. The horizontal axis symbolizes the control variable $\mathbf{f}$ on a line. The vertical axis shows the value of $\Delta U$, with negative values colored in blue. The cartoon shows four successive time steps (a)-(d). At each time step, the initial condition (orange) goes to a local minimum (red) which is used as initial condition for the following iteration. However, at step (e), in panel A, the local minimum is positive so an external reset is needed. Note how two different mechanisms at time step (e) in panels A and B produce the same effect. }
    \label{figS:figReset}
\end{figure}

The variation of the control variable $\mathbf{f}$ between two successive time steps is often small. However, large discontinuities may be occasionally present, in particular at the end of the protocol, see Task 1 and Fig.~\ref{fig:task0} in the main text. The fact that a discontinuous control can emerge in dynamical problems without discontinuities is well known, for instance in optimal control~\cite{lenhart2007optimal}. These discontinuities may have two origins. The first is an external reset, see step 2 in the loop above. If the control variable is stuck in a minimum with positive $\Delta U(\mathbf{f})$ (or higher than the threshold), the gradient descent is reset with a different initial condition, see Fig.~\ref{figS:figReset}A. This may also happen if, before the gradient descent starts, if the previous control $\mathbf{f}$ is worse than a randomly drawn new one. The second possibility is that the gradient descent of $W$ finds a path to access a new distant solution, which was previously inaccessible, see Fig.~\ref{figS:figReset}B.

The choice of hyper-parameters is the following: $q=1/10\,f_{max}$, $n_{av}=10$ (size of the sliding window for running average), $\alpha$ is chosen from $\alpha=10^{-s}\,\left\|\frac{dW}{d\mathbf{f}}\right\|^{-1}$ with $s\in\{-1+k/2|k=0,1,...,10\}$. A few hyper-parameters are task-specific. Task~1: $\Delta U^{threshold}=0$; $\Delta W^{threshold}=(\Delta U^{threshold}-W)/500$ if $\Delta U^{threshold}>W$, else $\Delta W^{threshold}=(\Delta U^{threshold}-W)/(500+50\,\min(n_{resets},15)-k)$ where $k$ is the iteration number after the last reset (or beginning of the loop) and $n_{reset}$ is the number of resets; when resetting $\mathbf{f}$, components $f_i$ are drawn independently and uniformly between $[f_{min},f_{max}]$. Task~2: $U^{threshold}=0.001/\Delta t$ if $U>0.7$ and 0 otherwise and $\Delta W^{threshold}$ as in Task 1; 
when resetting $\mathbf{f}$, components $f_i$ are drawn independently and uniformly between $[0,f_{max}]$.
Additional details can be found in the code.

\section{Further information on Task 1}\label{supp:task1}

\subsection{Effect of parameters}\label{supp:param}
Depending on the choice of parameters and hyper-parameters the training process can change from fast and easy to very hard or even impossible. We consider in particular the role of the size of the network, the control strength range and the choice of parameter in the learning rule. We stress that the hyper-parameters (including the time after which the training was interrupted) were optimized for the setting reported in the main text and were not changed as we varied these parameters. 

\begin{figure}[h!]
    \centering
     \includegraphics[width=.8\textwidth]{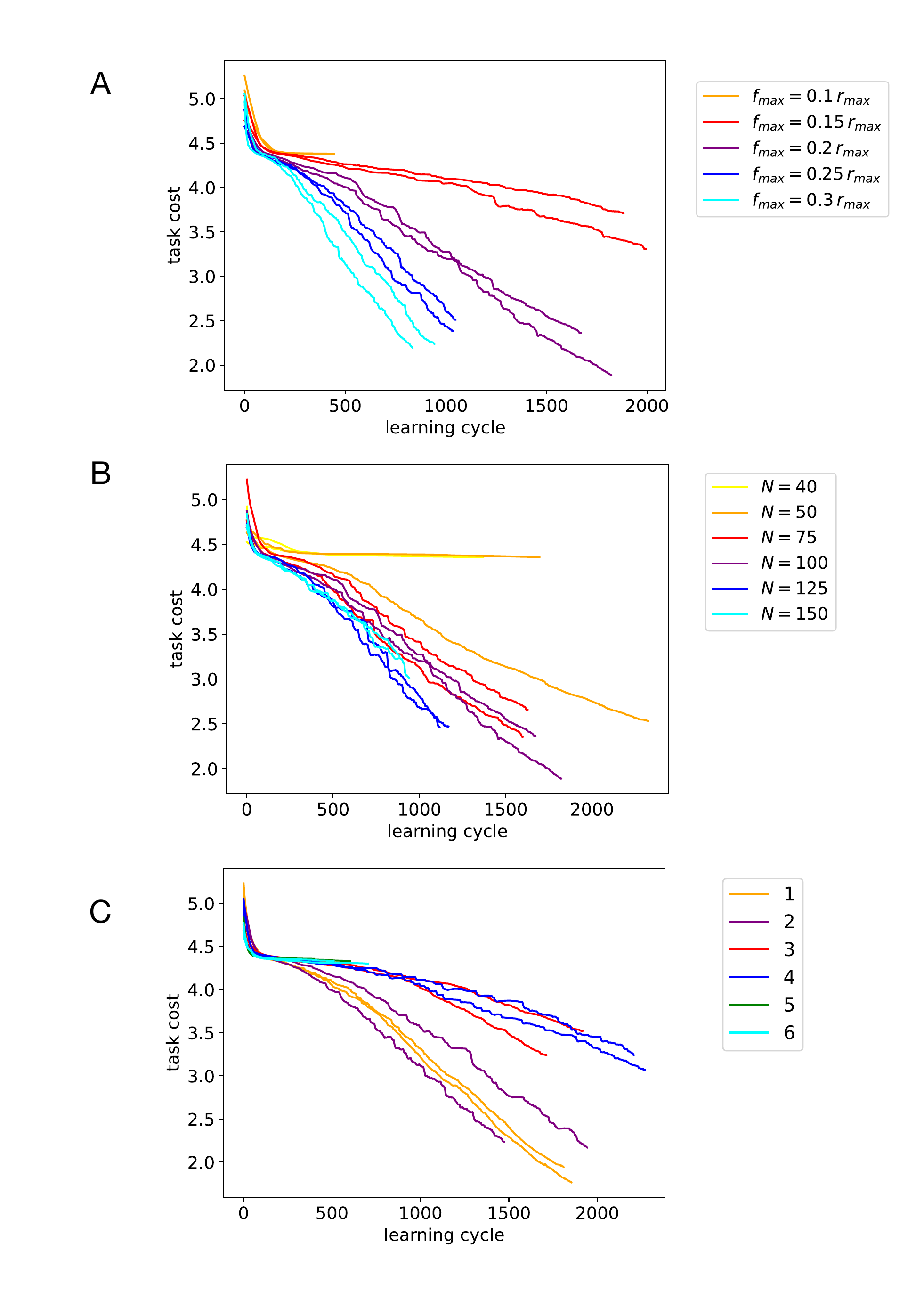}
    \caption {Training cost as a function of the learning cycle for Task 1  with parameter variations. Two examples, colored in the same way, are shown for each case. A: Changing the control maximal value $f_{max}$.  Notice that, for $f_{max}=0.1\,r_{max}$, training fails.  B: Changing the network size $N$; in practice, $N_{in}$ and $N_{out}$ are left unchanged, while $N_{proc}=N-30$ is varied. C: Changing other parameters, see descriptions of settings 1 to 6 in the text. }
    \label{figS:figF}
\end{figure}

In Fig.~\ref{figS:figF}A, we study the dependency of the training curves with the size $N$ of the network for $f_{max}=0.2\,r_{max}$ as in the main text. We observe that learning becomes harder (longer) and shows increased variability for smaller networks. Note the non monotonic behaviour in $N$, which should, however, be interpreted with caution since the parameter scaling with the size should be treated more carefully. 

We now consider the role of the control strength $f_{max}$ for networks with $N=100$ neurons. Intuitively, the larger the range, the easier the control. This is confirmed by Fig.~\ref{figS:figF}B, where the training curves are shown for five values of $f_{max}/r_{max}=0.1$, 0.15, 0.2, 0.25, 0.3; 0.2 is the value used in the main text). It is important to highlight that if the control is too weak, training seems to fail. While this minimal value of $f_{max}$ correspond to Task 1, and is not expected to hold for other tasks\footnote{In particular, it is likely that the value of the minimal $f_{max}$ could be slightly improved by optimizing over the hyper-parameters.}, we expect the existence of a critical value of $f_{max}$ below which learning through  stimulation cannot be implemented to be quite generic.

In Fig.~\ref{figS:figF}C,  we change some parameter values, focusing on $\beta_1$,$\theta(E)$ and $\theta(I)$. 
\begin{itemize}
    \item Setting 1: $\theta(E)=0.08\,r_{max}$, $\theta(I)=0.12\,r_{max}$, $\beta_1=4$,  $\theta_0(E)=\theta_0(I)=0.16\,r_{max}$; 
    \item Setting 2 (as in the main text): $\theta(E)=0.08\,r_{max}$, $\theta(I)=0.12\,r_{max}$, $\beta_1=0.8$,  $\theta_0(E)=\theta_0(I)=0.16\,r_{max}$;
    \item Setting 3: $\theta(E)=0.1\,r_{max}$, $\theta(I)=0.06\,r_{max}$, $\beta_1=4.$,  $\theta_0(E)=\theta_0(I)=0.16\,r_{max}$; 
    \item Setting 4: $\theta(E)=0.1\,r_{max}$, $\theta(I)=0.06\,r_{max}$ $\beta_1=0.8$,  $\theta_0(E)=\theta_0(I)=0.16\,r_{max}$;
    \item Setting 5: $\theta(E)=0.12\,r_{max}$, $\theta(I)=0.1\,r_{max}$, $\beta_1=4$,  $\theta_0(E)=\theta_0(I)=0.16\,r_{max}$; 
    \item Setting 6: $\theta(E)=0.12\,r_{max}$, $\theta(I)=0.1\,r_{max}$, $\beta_1=0.8$,  $\theta_0(E)=\theta_0(I)=0.16\,r_{max}$.
\end{itemize}   
All the other parameters keep the same values as in the main text. In particular all results in Fig.~\ref{figS:figF}C were obtained with $N=100$ neurons.

\subsection{Stability to uncertainty in the plasticity rules}\label{SI:error}

An important assumption in our work is that we have perfect knowledge of the plasticity rule to compute the optimal control protocol. To go beyond this idealized setting,  we hereafter study how mismatches between the plasticity rule used to make the network evolve and for computing the control affect the performances.

In practice, we consider two versions of the learning rule (Eq.~\ref{eqpl} in main text) corresponding to two different sets of parameters. The second rule is obtained from the first one upon insertion of multiplicative and independent errors, uniformly distributed between $1-$Err and $1+$Err, on $\theta(\epsilon_i)$, $\theta_0(\epsilon_i)$, $\eta(\epsilon_i)$ (independently for each synapse $i,j$) and on the parameters $\beta_1$ and $\beta_2$. The relative error Err takes the same value for all parameters.

\begin{figure}[h!]
    \centering
     \includegraphics[width=.6\textwidth]{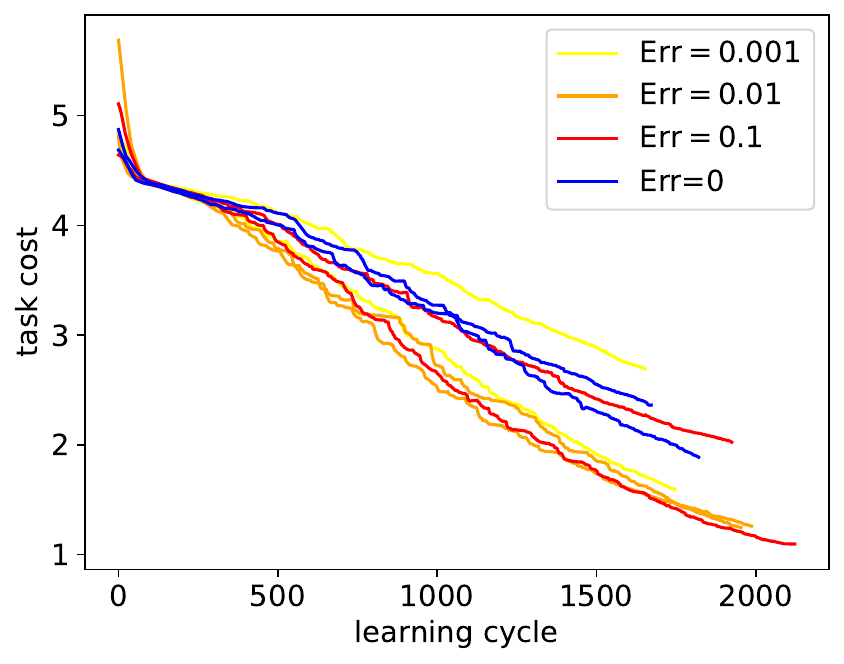}
    \caption {Evolution of the cost for Task 1 for different values of the errors on the learning rule parameters. Two trials are shown for each value of Err.
     }    \label{figS:figErr}
\end{figure}

We show in Fig.~\ref{figS:figErr} that a small but negligible error on the parameters of the learning rule does not significantly impact the performance in Task 1. More precisely, for moderate values of Err, the bias introduced by the mismatch between the 'true' and model learning rules remains smaller than the fluctuations in performance from trial to trial.

\color{black}

\subsection{Case of Hebbian rule for inhibitory neurons}\label{supp:ht3}

\begin{figure}[h!]
    \centering
     \includegraphics[width=.5\textwidth]{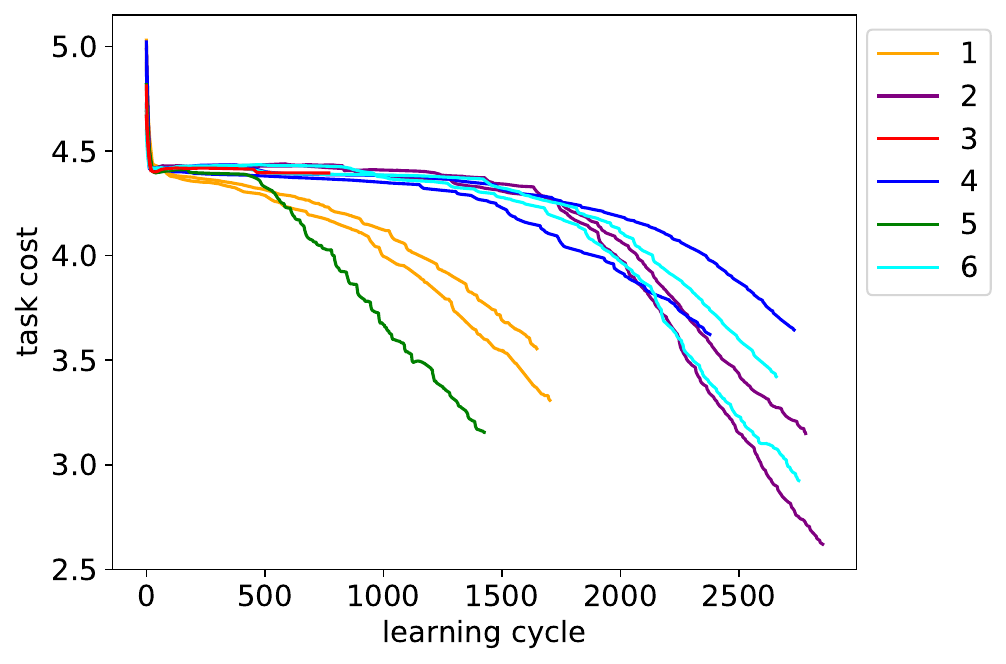}
    \caption {Hebbian rule for inhibitory neurons in Task 1. }
    \label{figS:figH}
\end{figure}

Last of all we show, in Fig.~\ref{figS:figH}, that Task 1 can also be done with Hebbian learning for inhibitory synapses. The picture is not much different from the anti-Hebbian case. We show a few settings with $N=100$. The settings are the following. 
\begin{itemize}
    \item Setting 1: $\theta(E)=0.2 r_{max}$, $\theta(I)=0.12 r_{max}$,  $\theta_0(E)=\theta_0(I)=0.08\,r_{max}$, $\beta_1=3.125$, $\beta_2=0.5\times (r_{max}/\bar J)^{2}$.
    \item Setting 2: $\theta(E)=0.06 r_{max}$, $\theta(I)=0.1 r_{max}$, $\theta_0(E)=\theta_0(I)=0.08\,r_{max}$, $\beta_1=3.125$, $\beta_2=0.5\times (r_{max}/\bar J)^{2}$.
    \item Setting 3: $\theta(E)=0.12 r_{max}$, $\theta(I)=0.08 r_{max}$, $\theta_0(E)=\theta_0(I)=0.08 r_{max}$, $\beta_1=3.125$, $\beta_2=0.5\times (r_{max}/\bar J)^{2}$.    
    \item Setting 4: $\theta(E)=0.08 r_{max}$, $\theta(I)=0.12 r_{max}$,  $\theta_0(E)=\theta_0(I)=0.08\,r_{max}$,$\beta_1=3.125$, $\beta_2=0.5\times (r_{max}/\bar J)^{2}$.  
    \item Setting 5: $\theta(E)=0.10 r_{max}$, $\theta(I)=0.06 r_{max}$, $\theta_0(E)=\theta_0(I)=0.08\,r_{max}$, $\beta_1=3.125$, $\beta_2=0.5\times (r_{max}/\bar J)^{2}$. 
    \item Setting 6: $\theta(E)=0.08 r_{max}$, $\theta(I)=0.12 r_{max}$, $\theta_0(E)=\theta_0(I)=0.08\,r_{max}$,$\beta_1=3.125$, $\beta_2=0.5\times (r_{max}/\bar J)^{2}$.
\end{itemize}

\subsection{Neural activation and input representation} 
\begin{figure}[h!]
    \centering
     \includegraphics[width=1.\textwidth]{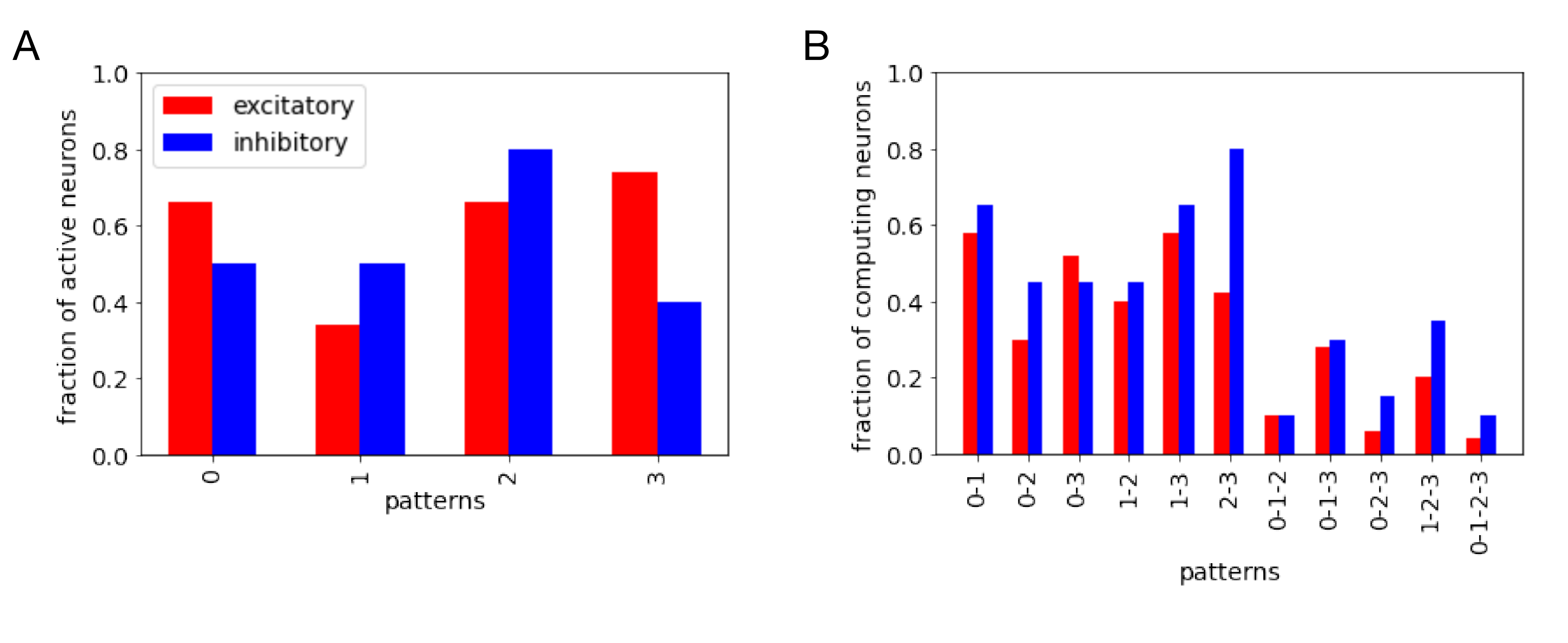}
    \caption {A: Bar plots of the fractions of active bulk neurons ($r/r_{max}>0.01$)  for each input. B:  fractions of neurons that separate the representations of two or more inputs. A neuron is said to separate two inputs $\mu$ and $\nu$ if its activity significantly varies with the applied input, {\em i.e.} $|r({\bf{f}}_\mu)-r({\bf{f}}_\nu)|/r_{max}>0.01$; this inequality should hold for any pair of inputs if more than two are considered.}
    \label{figS:figNC}
\end{figure}

If Fig.~\ref{figS:figNC} we provide a picture of how the neural network represents different inputs in the \emph{proc} region of the network and how the differentiation emerges. First we show the fraction of neuron activating (defined as being surpassing a certain activity threshold). We see in Fig.~\ref{figS:figNC}A that, no metter the input, a large fraction of neurons is always responding. In Fig.~\ref{figS:figNC}B we show the fraction of neurons which contribute in differentiating the representation of different digits. It is interesting to see that inhibitory neurons contribute the most.

\subsection{Correlation of activity along training}\label{supp:Corr}

\begin{figure}[h!]
    \centering
     \includegraphics[width=1.\textwidth]{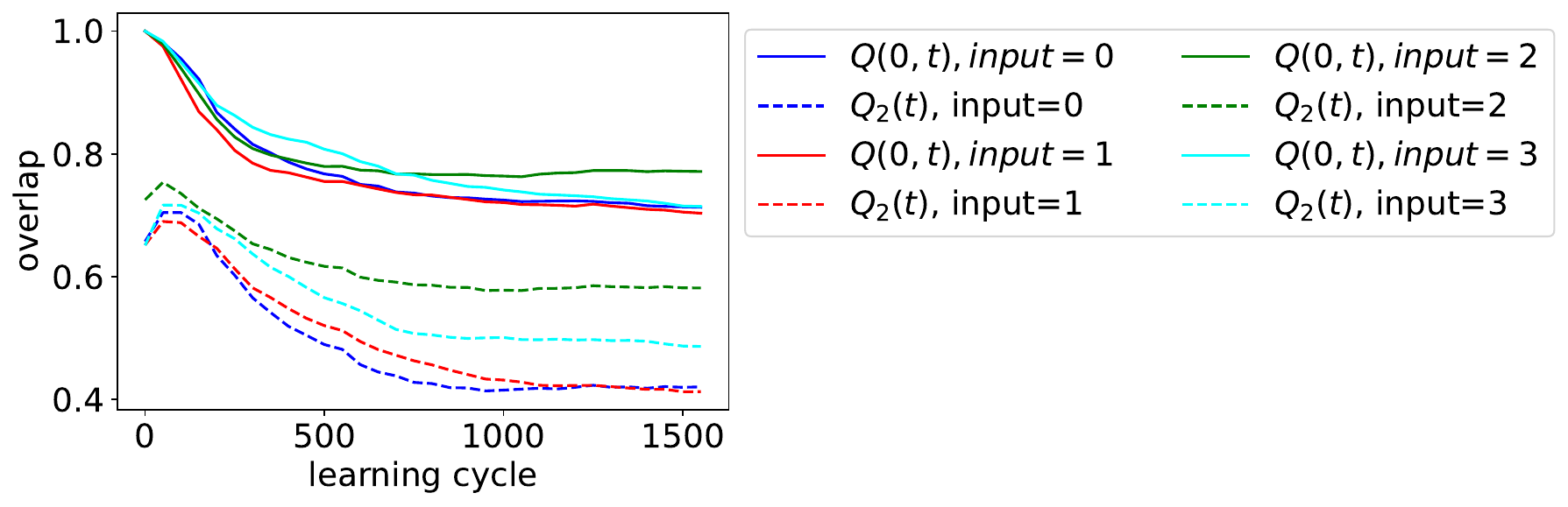}
    \caption {Overlaps $Q(0,t)$ and $Q_2(t)$ as functions of the learning cycle $t$ for each input pattern $\mu=0,1,2,3$  in Task 1. Results are shown after averaging over 10 trials. 
     }    \label{figS:figCorr}
\end{figure}

We investigate how the activity response to input patterns change across learning cycles. For each input pattern $\mu$, we compute the activity of the recurrent network area $\mathbf{r}_{rec}(k,q)$ as a function of the training cycle $k$ for each repetition $q$ of the training process ($n_{trial}=10$). 
We then define the trial-averaged overlap with the initial condition
\begin{equation}
    Q(0,k)=\frac{1}{n_{trials}}\sum_q\frac{\mathbf{r}_{rec}(k,q)\cdot\mathbf{r}_{rec}(0,q)}{\|\mathbf{r}_{rec}(k,q)\|\;\|\mathbf{r}_{rec}(0,q)\|}
\end{equation}
and the average overlap among trials
\begin{equation}
    Q_2(k)=\frac{2}{n_{trials}(n_{trials}-1)}\sum_{p>q,q=1}^{n_{trials}}\frac{\mathbf{r}_{rec}(k,q)\cdot\mathbf{r}_{rec}(k,q)}{\|\mathbf{r}_{rec}(k,p)\|\;\|\mathbf{r}_{rec}(k,p)\|}\ .
\end{equation}
Figure~\ref{figS:figCorr} shows that the activity remains highly correlated with its initial value and that different trials do not converge towards the same activity. This result can be interpreted as the presence of a strong dependence on the initial condition in the network at the end of the training, which is due the existence of multiple network solutions of the functional task.

\section{Further information on Task 2} \label{supp:task2}

\subsection{Multiple solutions for the stationary activity}

As the target connectivity in Figure~\ref{fig:fig5}A has its largest singular value larger than one, the stationary equations for the neural activity may have multiple solutions. Multiple solutions are rare but do occur occasionally (notice that limit cycles are in principle possible but were never observed during the stimulation protocol). 

When multiple solutions are present, it is possible that the control optimization loop selects a certain control stimulation pattern $\mathbf{f}^*$ associated to a specific state of neural activity $\mathbf{r}^*$ but, once the control stimulation is applied , another state of activity, say, $\mathbf{r}'$, is reached by the neural dynamics. We stress that we are addressing here the case of a large difference between $\mathbf{r}^*$ and $\mathbf{r}'$. Small differences are always expected to take place due to the small discrepancies between the true (and unknown) connectivity and its estimate used for the computation of the optimal control.
Since the plasticity inducing stimulation is long compared to $\tau_r$, we  assume it can be interrupted in time should this scenario take place. Then, the control stimulation is applied again. This goes on until the correct fixed point is selected. This procedure is effective only if there is a limited number of coexisting fixed points, as appears to be the case. In practice,  if the wrong solution is reached during the control stimulation phase, the fixed point iteration of the neural dynamics is restarted with a new initial condition.

\subsection{Choice of probing stimulations for connectivity inference}\label{supp:stimtask2}

While the activation function $\Phi$ in Eq.~(8) (Methods, main text) is invertible, $\Phi^{-1}$ is ill conditioned for very low firing rates. 
To avoid numerical issues in the estimation of the connectivity, we consider probing stimulations such that the least active neuron still has detectable activity. We use stronger stimulations for excitatory neurons than for inhibitory ones.  Excitatory neurons are stimulated with a random forcing in the range $[f_{max}/2,f_{max}]$, while for inhibitory neurons, the range considered is $[0,f_{max}/2]$. Despite this procedure, if a neuron  $j$ is found to have low activity (less than $0.2\; 10^{-3}\,r_{max}$), then its stimulation $f_j$ is increased by a random amount in the range $[0,f_{max}/2]$ (and the value is then cropped at $f_{max}$ if necessary). If, after correcting all the components $f_j$, any neuron remain inactive, the stimulation pattern is discarded and another one is randomly drawn. 

At the beginning of the training phase, this procedure is largely unnecessary, but at later (training) times, large quasi-disconnected ({\em i.e.} with very weak synaptic connections) groups of neurons appear. At this point, the pattern correction procedure becomes essential. After some time, all neurons will be again connected together and  corrections are less and less frequent. We stress that correcting a probing stimulation pattern does not require to stimulate the system more times as,  based on our relatively accurate estimate of the network connectivity, we can predict the response of the network in a reliable way.

\subsection{Case of anti-Hebbian rule for inhibitory neurons}\label{aht2}

\begin{figure}[h!]
    \centering
     \includegraphics[width=1.\textwidth]{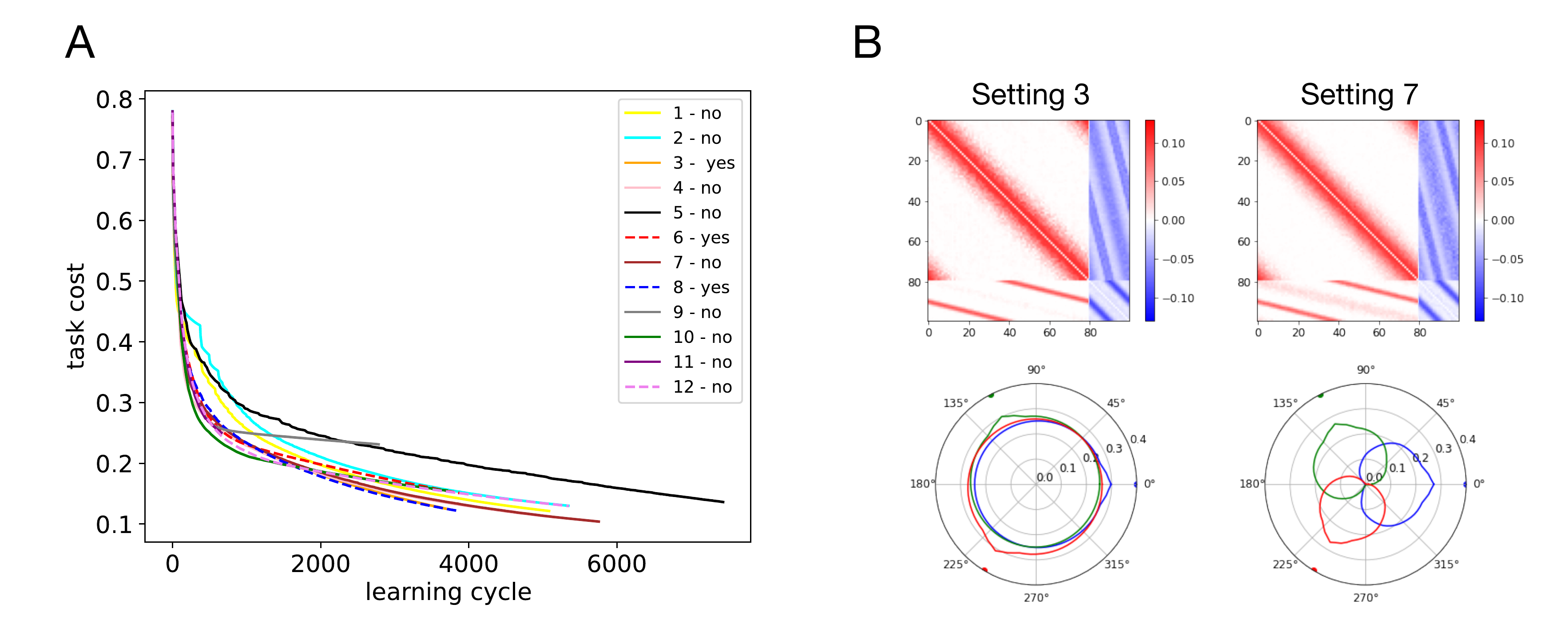}
    \caption {Panel A: anti-Hebbian inhibitory synapses. We show different learning curves for 12 different parameter settings for Task 2. The label on each curve (yes/no) tells if the attractor was generated or not. Dotted lines are used to make overlapping lines visible. Panel B shows the connectivity matrix $\mathbf{J} $ and the receptive fields for settings 3 and 7 at the end of the training. Notice how similar connectivity structures and cost values are associated to very different performances in terms of receptive field formation.}
   \label{figS:fig2anti}
\end{figure}

In this section we train our network on Task 2  with anti-Hebbian learning rule for inhibitory synapses (see~\ref{figS:fig2anti}). We show twelve different settings, highlighting a certain variability in the learning curves.
Indeed, the protocol works as intended in the sense that the structural cost decreases. However, if we look at the continuous attractor behaviour, we get a different picture. For instance, we can focus on setting 3 and 7, for which the final cost is approximately the same at the end of the protocol. Likewise, the connectivity structures are very similar (see Fig.~\ref{figS:fig2anti}B). However, in the setting 7, receptive fields appear, while in setting 3 they do not, since. The reason is that the final result is not given by a random but unbiased perturbation of the target connectivity state but, rather, a biased version of it, due to the effect of plasticity constraints in the learning process. Specifically, we can clearly see the interference of the formation of different ``stripes'' in the connectivity matrix. 

The settings are the following (we only specify parameters differing from those of the main text). 
\begin{itemize}
    \item $\theta(E)=0.175\,r_{max}$, $\theta(I)=0.1\,r_{max}$, $\eta(I)=-1.1$, $\beta_1=6.25$, $\beta_2=2.5\times (r_{max}/\bar J)^2$,  $\theta_0(E)=\theta_0(I)=0.16\,r_{max}$
    \item $\theta(E)=7\,r_{max}/30$, $\theta(I)=4\,r_{max}/30$, $\eta(I)=-1$, $\beta_1=6.25$, $\beta_2=2.5\times (r_{max}/\bar J)^2$,  $\theta_0(E)=\theta_0(I)=0.16\,r_{max}$
    \item $\theta(E)=6\,r_{max}/70$, $\theta(I)=6\,r_{max}/70$, $\eta(I)=-1.5$, $\beta_1=6.25$, $\beta_2=2.5\times (r_{max}/\bar J)^2$,  $\theta_0(E)=\theta_0(I)=0.16\,r_{max}$ 
    \item $\theta(E)=4\,r_{max}/35$, $\theta(I)=2\,r_{max}/35$, $\eta(I)=-0.9$, $\beta_1=3.125$, $\beta_2=2.5\times (r_{max}/\bar J)^2$,  $\theta_0(E)=\theta_0(I)=0.16\,r_{max}$
    \item $\theta(E)=2\,r_{max}/35$, $\theta(I)=3\,r_{max}/35$, $\eta(I)=-1.5$, $\beta_1=1.25$, $\beta_2=2.5\times (r_{max}/\bar J)^2$,  $\theta_0(E)=\theta_0(I)=0.16\,r_{max}$
    \item $\theta(E)=0.12\,r_{max}$, $\theta(I)=0.08\,r_{max}$, $\eta(I)=-2.$, $\beta_1=3.125$, $\beta_2=2.5\times (r_{max}/\bar J)^2$,  $\theta_0(E)=\theta_0(I)=0.16\,r_{max}$
    \item  $\theta(E)=0.16\,r_{max}$, $\theta(I)=0.08\,r_{max}$, $\eta(I)=-1.1$, $\beta_1=3.125$, $\beta_2=2.5\times (r_{max}/\bar J)^2$,  $\theta_0(E)=\theta_0(I)=0.16\,r_{max}$
    \item  $\theta(E)=0.16\,r_{max}$, $\theta(I)=0.12\,r_{max}$, $\eta(I)=-1.1$, $\beta_1=3.125$, $\beta_2=2.5\times (r_{max}/\bar J)^2$,  $\theta_0(E)=\theta_0(I)=0.16\,r_{max}$
    \item $\theta(E)=0.16\,r_{max}$, $\theta(I)=0.12\,r_{max}$, $\eta(I)=-3.$, $\beta_1=3.125$, $\beta_2=2.5\times (r_{max}/\bar J)^2$,  $\theta_0(E)=\theta_0(I)=0.16\,r_{max}$
    \item $\theta(E)=0.16\,r_{max}$, $\theta(I)=0.12\,r_{max}$, $\eta(I)=-1.3$, $\beta_1=1.25$, $\beta_2=2.5\times (r_{max}/\bar J)^2$,  $\theta_0(E)=\theta_0(I)=0.16\,r_{max}$
    \item $\theta(E)=0.12\,r_{max}$, $\theta(I)=0.12\,r_{max}$, $\eta(I)=-1.3$, $\beta_1=3.125$, $\beta_2=2.5\times (r_{max}/\bar J)^2$,  $\theta_0(E)=\theta_0(I)=0.16\,r_{max}$
    \item $\theta(E)=0.12\,r_{max}$, $\theta(I)=0.08\,r_{max}$, $\eta(I)=-1.3$, $\beta_1=1.25$, $\beta_2=2.5\times (r_{max}/\bar J)^2$,  $\theta_0(E)=\theta_0(I)=0.16\,r_{max}$
\end{itemize}

\subsection{Effect of variation of parameters}
\begin{figure}[h!]
    \centering
     \includegraphics[width=.5\textwidth]{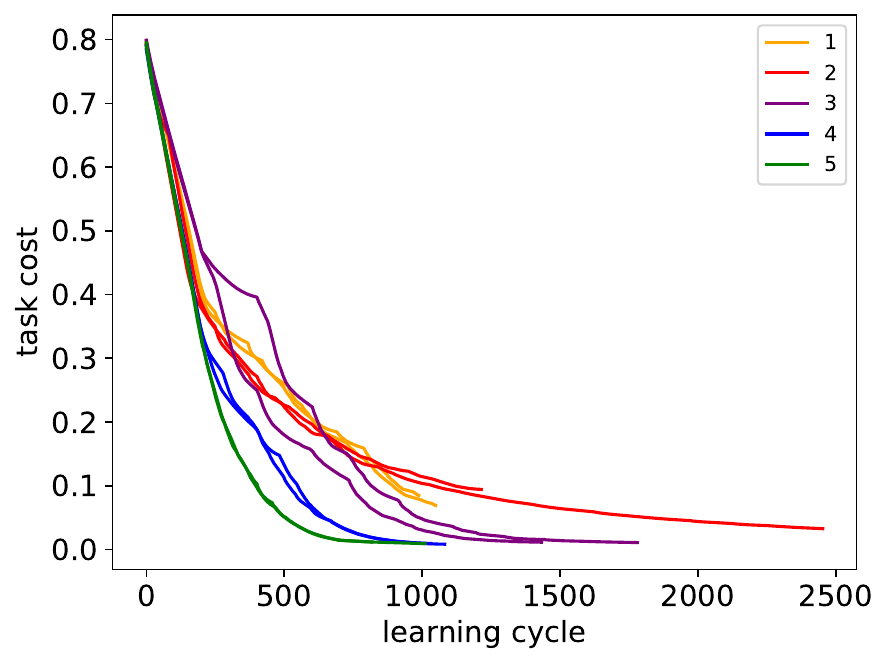}
    \caption {
     Learning curves for Task~2 with Hebbian inhibitory connections (as in the main text) but with different parameter choices.  }    \label{figS:figCan_2}
\end{figure}

In Fig.~\ref{figS:figCan_2} we show the learning curves associated to different sets of parameters. We only mention parameters differing from the setting in the main text and, in particular, we provide some example of the effect of perturbation of the hebbian post-synaptic threshold
\begin{itemize}
    \item $\theta(E)=0.12\,r_{max}$, $\theta(I)=0.08\,r_{max}$ ,  $\theta_0(E)=\theta_0(I)=0.08\,r_{max}$
    \item $\theta(E)=0.08\,r_{max}$, $\theta(I)=0.12\,r_{max}$,  $\theta_0(E)=\theta_0(I)=0.08\,r_{max}$
    \item 
    $\theta(E)=0.2\,r_{max}$, $\theta(I)=0.16\,r_{max}$,  $\theta_0(E)=\theta_0(I)=0.08\,r_{max}$ (as in the main text) 
    \item $\theta(E)=0.2\,r_{max}$, $\theta(I)=0.08\,r_{max}$ ,  $\theta_0(E)=\theta_0(I)=0.08\,r_{max}$
    \item $\theta(E)=0.16\,r_{max}$, $\theta(I)=0.12\,r_{max}$ ,  $\theta_0(E)=\theta_0(I)=0.08\,r_{max}$
\end{itemize}

\section{Additional Task 3: logical AND function in a small circuit}~\label{supp:task0}

\begin{figure}[h!]
    \centering
         \includegraphics[width=.9\textwidth]{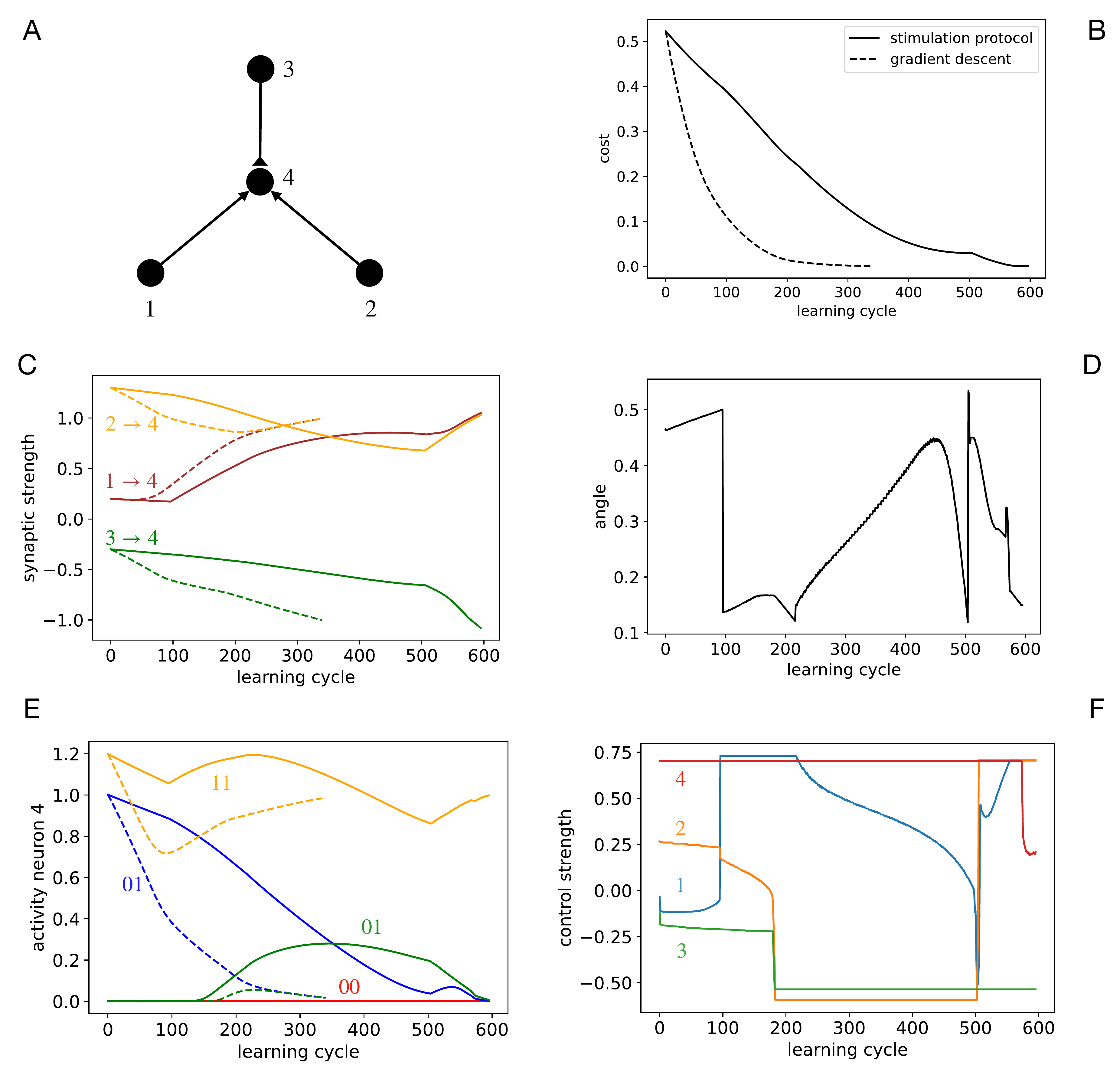}
    \caption{  
{A} Example of a 4-neuron circuit implementing a logical AND between two inputs. Two excitatory (nodes 1 and 2) and one inhibitory (node 3) neurons are connected to an output neuron (node 4). The three non zero connections are approximately $J_{41}=1$, $J_{42}=1$ and $J_{43}=-1$ at the end of the training. We set $f_{max}=0.7\,\bar r$ and $f_{min}=-0.5\,\bar r$.
{ B}: Cost $U_{task}$ as a function of the time step (time $t/\Delta t$) during learning with our stimulation protocol (full curve) and standard gradient descent (dashed line). For the latter, we choose the learning rate 
in such a way that the first connectivity change  has the same norm $\|\Delta\mathbf{ J}\|$  as with the stimulation protocol. 
{C}: activity $r_4$ of neuron 4 (in $\bar r$ units)in response to the four possible pairs of inputs onto neurons 1 and 2 through the training process.
{D}: Cosine similarity between (minus) the gradient of the cost $\partial U/\partial \mathbf{J}$ and the time derivative of the connectivity $\dot {\mathbf{J}}$. The gap between the similarity and one (corresponding to gradient descent) reflects the limited accessibility of the connectivity space imposed by the plasticity rule, see Fig.~\ref{fig:fig3} from the main text. Notice that the similarity remains positive at all times, implying that the plasticity rule can be exploited to decrease the cost and learn the task.
{ (E)} Evolution of the synaptic strengths during learning with our stimulation protocol (full lines) and standard gradient descent (dashed lines).
{ (F)} control $f_i(t)$ (in $\bar r$ units) applied to the neurons $i=1,2,3,4$. The control stimulations smoothly vary for a certain period of time, then abruptly jump when no local solution for the control optimization process are found near that at previous step.      }
    \label{fig:task0}
\end{figure}

\paragraph{Definition of the task.} We propose an additional simple task in a feed-forward network, with few neurons and simplified activation function, which lacks the complexity and theoretical interest of the tasks in the main text, but offers a detailed visualization of the quantities involved in learning. We set $r_{max}=\infty$ and $r_0=0$ (i.e. we use the standard , i.e. the standard ReLU as activation function). Hence, we cannot express quantities in units of $r_{max}$. Instead, we use as a unit of measure a value $\bar r$ which is assumed to be an intermediate frequency $r_0\ll \bar r \ll r_{max}$. The task consists in the realization of a logical AND between two inputs in a circuit made of $N=4$ neurons, out which $N_E=3$ are excitatory and $N_I=1$ is inhibitory. We assume neurons 1 and 2 can receive binary input stimulations, in practice $f_1,f_2=0\times\bar r$ or $1\times\bar r$ (which we identify, respectively, with logical values False=`0' and True=`1'), while we set $f_3=1 \,\bar r$.  We demand that the activity $r_4$ of neuron 4 expresses the logical AND, i.e. $r_4=\bar r(=`1')$ if both $f_1$ and $f_2$ are `1', and $r_4=0(=`0')$  otherwise.
A possible circuit implementing this task is shown in Fig.~\ref{fig:task0}. Two excitatory neurons (nodes 1 and 2) and an interneuron (node 3) are connected to an output neuron (node 4). The non-zero connections are set to $J_{41}=J_{42}=-J_{43}=1$:  
the inhibitory input to neuron 4 coming from neuron 3 is large enough to suppress the inputs from either neuron 1 or 2, but not from both. 

We next train a network with random initial connectivity with the following cost, see Eq.~\eqref{cost:functional},
\begin{equation}\label{cost:task1}
    U_{task} \big(\mathbf{J} \big)=\frac 14\bigg[ \big(r_4(1,1)-\bar r\big)^2+\big(r_4(0,1)-0\big)^2+\big(r_4(1,0)-0\big)^2+\big(r_4(0,0)-0\big)^2 \bigg]\ ,
\end{equation}
which is the sum of squared errors for the four possible combinations of binary inputs. The activities of neurons 1, 2 and 3 are not considered in the cost. This cost is functional and not structural: it does not impose {\em a priori} any target value for the connectivity matrix, but rather constrains the function to the achieved. 

\paragraph{Training: stimulations and network.}
Despite its simplicity, the AND task case highlights the underlying difficulty in training a plastic network and the difference with the standard gradient descent (GD) used in machine learning. Figure~\ref{fig:task0}B shows the cost dynamics with GD and with our stimulation protocol. While both algorithms successfully learn the task, GD is faster, as we expect. For the plastic network, the three synapses cannot be updated independently, since the activity of the post-synaptic neuron 4, which enters both the Hebbian and the homeostatic part of the plasticity rule, influences at once the evolution of all the connections we are trying to update. As a result of this constraint, the changes in the $\mathbf{J}$ matrix cannot align along the gradient of the cost as is the case for the GD dynamics, see Fig.~\ref{fig:task0}D. Our protocol is nevertheless able to make the cost decrease at all times. This decrease demonstrated that the task is progressively learned, as visible from the responses of neuron 4 to all possible pairs of inputs reported in Fig.~\ref{fig:task0}E. 

The behaviour of the connectivity is displayed in Fig.~\ref{fig:task0}C. We observe that our stimulation protocols drives the network connectivity to the circuit shown in Fig.~\ref{fig:task0}A. It is easy to check that this network is the only one achieving zero cost, see Eq.~\eqref{cost:task1} and Methods. For larger networks, we expect the task to be achieved by many connectivity networks \cite{Marder}, and our learning protocol will select one of them.

The time course of the applied stimulations $f_i(t)$ is displayed in Fig.~\ref{fig:task0}F. We observe that, most of the time, the control smoothly changes across stimulation periods, but occasionally shows large discontinuities (as explained in Fig.~\ref{figS:figReset}) in this case, no discontinuity is due to an external reset. 

\paragraph*{Some remarks about the solution.}
Assuming symmetry between neurons 1 and 2, we can call $J_+=J_{41}=J_{42}$ and $J_-=-J_{43}$. Then, essentially, the task is solved when we have $1=\max(2J_+-J_-,0)=1$ and $\max(J_+-J_-,0)=0$ (and $\max(-J_-,0)=0$ which is always true). It means that the solution is given by the curve $J_-=2J_+-1$ in the $\{J_-\geq J_+\}\cap\{J_\pm>0\}$ space, which is the $J_-=2J_+-1$ curve for $J_+\geq 1$. Since we start with small synaptic strength values, it is not surprising the smallest possible values for the connections are selected by optimization procedure, as shown in Fig.~\ref{fig:task0}C.
\paragraph*{Parameters.}
We use $\Delta t=10^{-2}\tau_n$, $\theta(E)=\theta(I)=\theta_0(E)=\theta_0(I)=\bar r/2$, $\beta_1=0.2$, $\beta_2=0$.

%

\section{Description of videos}

\subsection{network\_formation.gif}\label{SI:gif1}

The video shows how,  upon  training, a sub-network is activated and sustains a directional flow of currents from the input region towards the output one. To identify this sub-network, we remove nodes that are inactive, in the sense that they receive or generate a current $I_i=\sum_j J_{ij} r_j$ smaller than a fixed (and arbitrary) threshold. More precisely, node $i$ is retained in the visualization if $\max \left(I_i,\left|\sum_j J_{ij}\right|r_i\right)>Th$, where $Th$ is the minimum threshold for which all active neurons meant to be active in the output digit have firing rates larger than $.18 r_{\max} $ at cycle $k=1500$ of the training procedure. Neurons fullifing this criterion are drawn and are connected by a directional arrow $i\to j$ if the contribution   $c_{ji}=J_{ij}r_j$ to the current $I_i$ is larger than $0.005 \,r_{\max}$. The input area is located on the left side of the network, and externally stimulated neurons are shown in red. Neurons in the recurrent part are represented by empty black circles, and output neurons are located on the right side. Frames are 50 cycle apart from each others. 

\subsection{activity\_patterns.gif }\label{SI:gif2}
Frames represent successive stages of learning in Task 1 and are separated by 50 learning cycles each. The network is divided into an input (IN), output (OUT) and bulk/recurrent (BULK) parts.
\begin{itemize}
    \item Top left panel: distributions of the four different types of connections (E/I $\to$ E/I). Large value of the interactions become more common as learning proceeds, but $I\to I$ connections are suppressed. 
    \item Top right panel: spatio-temporal pattern of stimulation shown with the same color coding as in the main text; blue and red represent, respectively, $f_i=-0.2r_{max}$ and $f_i=0.2r_{max}$. 
    \item Bottom panels:  activity of the network when stimulated in the IN area with each one of the four input patterns (in red). White denotes no activity, while black corresponds to $r\geq 0.04\,r_{\max}$.  Patterns of activity remain correlated along training, implying that the final solution is strongly determined by the initial condition.
\end{itemize}

\end{document}